\documentclass[%
reprint,
showpacs,reprintnumbers,
 amsmath,amssymb,
 aps,
prl,
superscriptaddress,
showpacs, showkeys]{revtex4-1}
\usepackage{amsmath}
\usepackage{subfigure}
\usepackage{textgreek}
\usepackage{soul} 
\usepackage{color}
\usepackage{breqn}
\usepackage{graphicx}
\usepackage{dcolumn}
\usepackage{bm}
\usepackage{hyperref}
\usepackage{textcomp}
\usepackage{xr}
\hypersetup{bookmarksnumbered, pdfpagemode=UseOutlines, 
colorlinks=true, citecolor=blue, filecolor=blue, linkcolor=blue, urlcolor=blue}

 \makeatletter
 \let\cat@comma@active\@empty
 \makeatother

\graphicspath{/media/sid/Dropbox/2nd_paper
}

\begin{document}


\title{Spin relaxation 1/f noise in graphene}

\author{S. Omar} 
\thanks{corresponding author}
\email{s.omar@rug.nl}
\affiliation{The Zernike Institute for Advanced Materials University of Groningen Nijenborgh 4 9747 AG, Groningen, The Netherlands}
\author{M.H.D. Guimar\~{a}es }
\affiliation{The Zernike Institute for Advanced Materials University of Groningen Nijenborgh 4 9747 AG, Groningen, The Netherlands}
\affiliation{Kavli Institute at Cornell for Nanoscale Science Cornell University, Ithaca, NY – 14853, USA}
\author{A. Kaverzin}
\affiliation{The Zernike Institute for Advanced Materials University of Groningen Nijenborgh 4 9747 AG, Groningen, The Netherlands}

\author{B.J. van Wees}
\affiliation{The Zernike Institute for Advanced Materials University of Groningen Nijenborgh 4 9747 AG, Groningen, The Netherlands}%

\author{I.J. Vera-Marun}
\thanks{corresponding author}
\email{i.j.vera.marun@rug.nl}
\affiliation{The Zernike Institute for Advanced Materials University of Groningen Nijenborgh 4 9747 AG, Groningen, The Netherlands}
\affiliation{School of Physics and Astronomy The University of Manchester, Manchester M13 9PL, UK}
\date{\today}

\begin{abstract}
We report the first measurement of 1/f type noise associated with electronic spin transport,  using single layer graphene as a prototypical material with a large and tunable Hooge parameter. We identify the  presence of two contributions to the measured spin-dependent noise: contact polarization noise from the ferromagnetic electrodes, which can be filtered out using the cross-correlation method, and the noise originated from the spin relaxation processes. The noise magnitude for spin and charge transport differs by three orders of magnitude, implying different scattering mechanisms for the 1/f fluctuations in the charge and spin transport processes.  A modulation of the spin-dependent noise magnitude by changing the spin relaxation length and time indicates that the spin-flip processes dominate  the spin-dependent noise. 
 
\begin{description}
\item[PACS numbers]
 \verb+85.75.-d+, \verb+73.22.Pr+, \verb+75.76.j+
\end{description}
\end{abstract}

\maketitle
Noise in electronic transport is often treated as nuisance. However, it can have much more information than the average (mean) of the signal and  can probe the system dynamics in greater detail than conventional DC measurements \cite{landauer_condensed-matter_1998}. Low frequency fluctuations with a power spectral density (PSD) that depend inversely on frequency, also known as 1/f noise are commonly observed phenomena in solid state devices. 
A textbook explanation of the processes generating 1/f noise is given by the McWhorter model where traps are distributed over an energy range, leading to a distribution of characteristic time scales of trapping-detrapping processes of the electrons in the transport channel and causing slow fluctuations in conductivity \cite{jayaraman_1/f_1989, hooge_experimental_1981,dutta_low-frequency_1981}. 

 Graphene is an ideal material for spin transport due to low spin-orbit coupling and small hyperfine interactions \cite{ertler_electron_2009, huertas-hernando_spin-orbit-mediated_2009}. However, the experimentally observed spin relaxation time  $\tau_{\text{s}}$ $\sim$ 3 ns and spin relaxation length $\lambda_{\text{s}}$ $\sim$ 24 \textmu m are \cite{pep_2015} lower than the theoretically predicted  $\tau_{\text{s}} \sim$ 100 ns  and $\lambda_{\text{s}} \sim$ 100 \textmu m  \cite{branas_graphene, macdonald_graphene}. There are a number of experiments and theories suggesting that the charge and magnetic impurities present in graphene might play an important role for the lower value of observed spin relaxation time \cite{fabian_relaxation_sub,Folk_relaxation,fabian_resonant_scattering,review_Roche,omar_spin_2015}. It is an open question whether these impurities affect the spin transport in a similar way as the charge transport, or if the scattering mechanisms in both processes behave differently. For electronic transport in graphene, the effect of impurities can be studied via 1/f noise measurements. In a similar line, measuring low frequency fluctuations of the spin accumulation can unravel the role of impurities on the spin transport. 

In this work, we report for the first time observation of spin-dependent 1/f noise, which we study on graphene spin valves performed in a non-local geometry. We find that the extracted noise magnitude ($\gamma^{\text{s}}$) for the spin transport is three orders of magnitude higher  than the noise magnitude ($\gamma^{\text{c}}$) obtained from the local charge noise measurements, indicating different scattering mechanisms producing 1/f fluctuations in the charge and spin transport. Such a large difference had not been pointed out until now, although different scattering mechanisms for spin transport have been proposed before \cite{fabian_resonant_scattering, review_Roche}.
In a recent experiment, Arakawa $\textit{et al.}$ \cite{arakawa_shot_2015} measure a spin dependent shot noise due to the spin-injection process. Also, they rule out the effect of spin-flip scattering due to similar Fano factor values obtained for the charge and spin transport. In contrast, we measure the spin dependent noise in a different frequency regime and find out that the dominant scattering mechanisms contributing to the 1/f noise are the processes which flip the spins, giving rise to a higher noise magnitude compared to the charge transport and highlighting the role of impurities in the spin relaxation. 

In order to perform the spin-dependent noise measurements, we prepare graphene spin valves. Single layer graphene is contacted with 35 nm thick ferromagnetic cobalt electrodes with $\sim$ 0.8 nm thick TiO$_2$ tunnel barrier inserted in between for efficient spin injection and detection (see supplementary for fabrication details) \cite{Tombros_nature, omar_spin_2015}. We characterize two different regions of our sample. They are labeled as device A and device B for further discussion.
\begin{figure}
\includegraphics[scale=1]{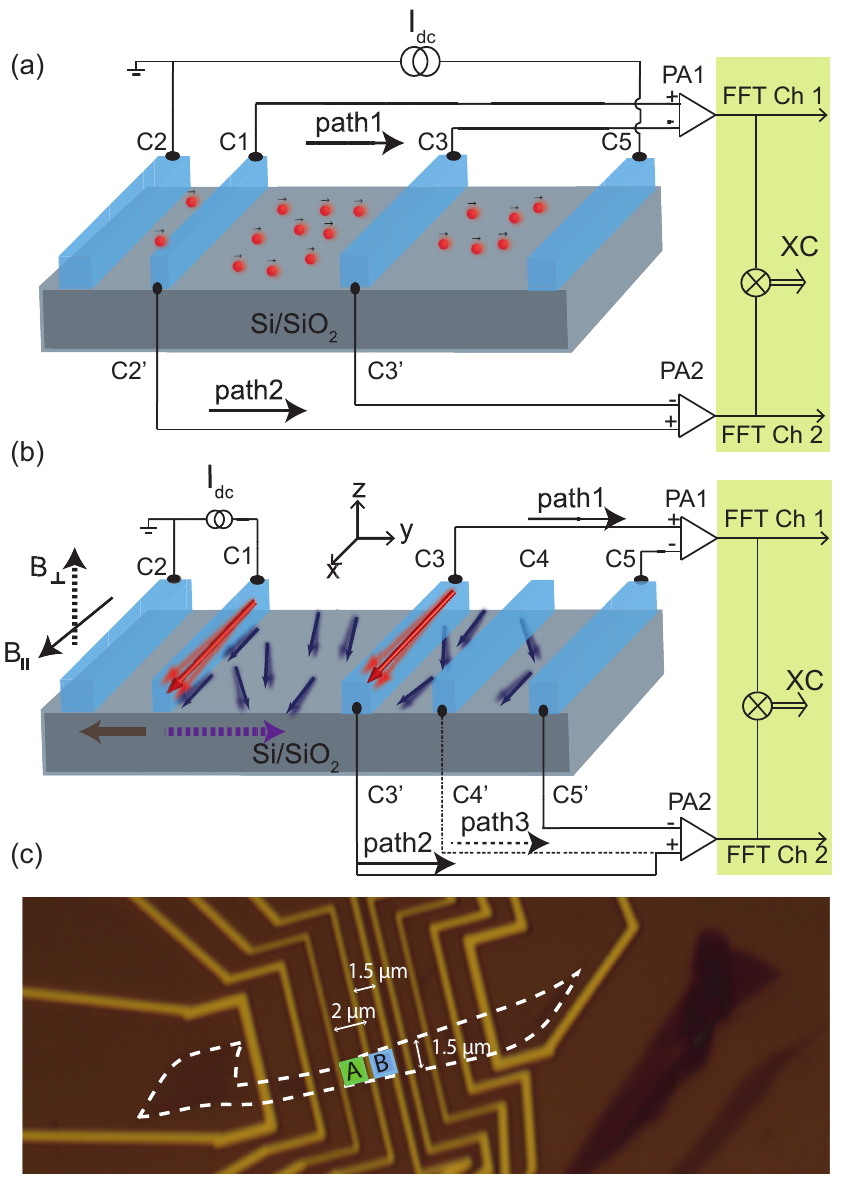}
\caption{\label{fig: measurement scheme}
(a) Cross correlation (XC) connection scheme for local charge noise measurement and (b) non-local spin-dependent noise measurements. A connection scheme for spatial cross correlation (SXC) is also shown  where the XC analysis is performed over the voltage measured between contacts C3-C5 ($V_{\text{NL}}^{C3-C5}$, path1) and contacts C4-C5  ($V_{\text{NL}}^{C4-C5(C4'-C5')}$, path 3). (c)  An optical picture of the sample of single layer graphene (white dotted line) connected via FM electrodes. 
Noise measurements are done in two regions of the sample, labeled A (l=2 $\mu$m, w=1.5 $\mu$m) and B (l=1.5 $\mu$m, w $\sim$ 1.5 $\mu$m). 
} 
\end{figure}
A lock-in detection technique is used for characterizing the charge and the spin transport properties.  All the measurements are carried out in high vacuum ($\sim$ 1 $\times$ 10$^{-7}$ mbar) at room temperature. For charge transport measurements we use the four probe connection scheme shown in Fig.~\ref{fig: measurement scheme}(a), which minimizes the contribution of the contacts. 

\begin{figure}
\includegraphics{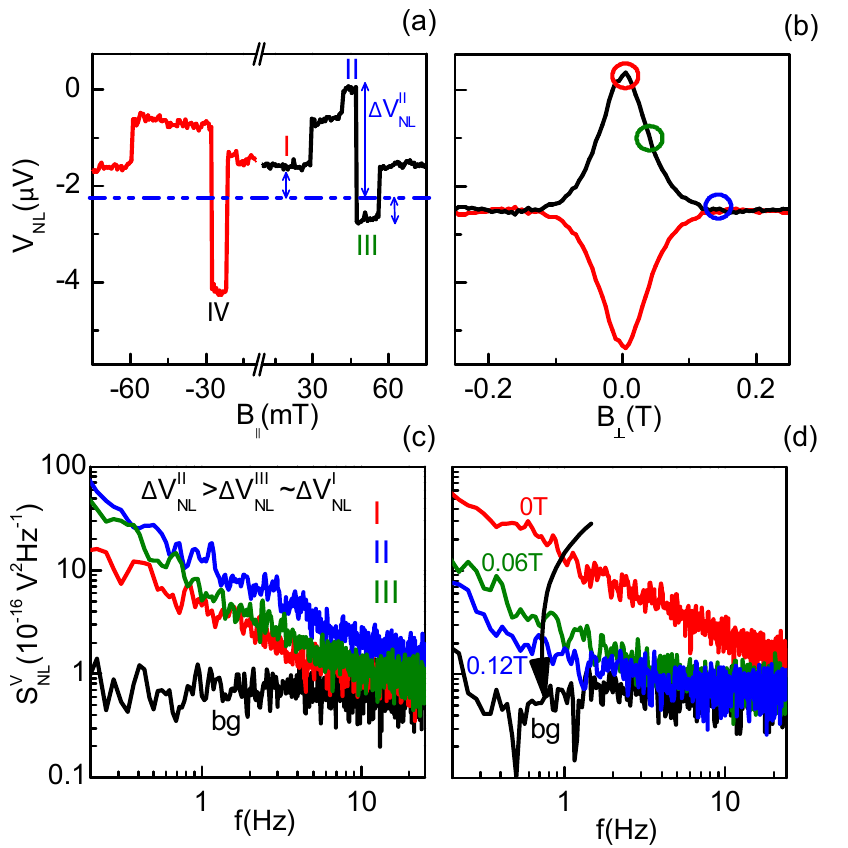}
 \caption{\label{fig:spin valve hanle noise}
(a) Non-local spin valve measurement. The dotted line represents the background level, which is estimated from the Hanle measurement. $\triangle V_{\text{NL}}$ is defined as the spin accumulation above or below with respect to the background level (see supplementary for switching details).  (b) Hanle measurement is shown for the level II and level IV of the spin valve. (c) noise PSD measured for the magnetization configurations corresponding to level I, II and III in the spin-valve measurement in Fig.~\ref{fig:spin valve hanle noise}(a). (d) the PSD plot for the Hanle configuration obtained at different magnetic fields, corresponding to the circles indicated in Fig.~\ref{fig:spin valve hanle noise}(b). In Fig.~\ref{fig:spin valve hanle noise}(c) and (d), \textasciigrave bg\textasciiacute represents the zero current background thermal noise.} 
 \end{figure}

 Spin transport is measured by applying a current between contacts C1-C2 to inject the spins into graphene and measure the spin accumulation between contacts C3-C5 (or C4-C5) in a four probe non-local detection scheme as shown in Fig.~\ref{fig: measurement scheme}(b). This method decouples the paths of the spin and charge transport and thus minimizes the contribution of the charge signal to the measured spin signal \cite{Tombros_nature}. In order to perform spin valve measurements, we first apply an in-plane high magnetic field ($B_{\parallel}$) along the easy axes of the ferromagnets to set their relative magnetization in the same direction. Then, the magnetic field is swept in the opposite direction in order to reverse the magnetization direction of the electrodes one by one depending on their coercivity. Each magnetization reversal appears as a sharp transition in the signal (Fig.~\ref{fig:spin valve hanle noise}(a)). For Hanle precession measurements, an out of plane magnetic field ($B_{\perp}$) is applied to precess the injected spins around the applied field for a fixed magnetization configuration of the ferromagnetic electrodes. A representative Hanle measurement from device A is shown in Fig.~\ref{fig:spin valve hanle noise}(b). With this measurement, we can extract the spin diffusion coefficient $D_{\text{s}}$ and spin relaxation time $\tau_{\text{s}}$, following the procedure described in ref.~\cite{Tombros_nature} and use them to calculate the contact polarization ($P$). For device A, we obtain $D_{\text{s}}\sim $0.03 m$^2$/s, $\tau_{\text{s}} \sim$ 110 ps and $P \sim$ 5\% and for device B, $D_{\text{s}} \sim$ 0.01 m$^2$/s, $\tau_{\text{s}} \sim$ 290 ps and $P \sim$ 10\%.

 In order to measure the noise from the sample, we use a two channel dynamic signal analyzer from Stanford Research System (model SR785) which  acquires the signal fluctuations in time  and converts it into a frequency domain signal via Fast Fourier Transform (FFT) algorithm. 

The 1/f noise of the charge transport in graphene is measured in a local four probe scheme, similar to the charge transport measurements (Fig.~\ref{fig: measurement scheme}(a)). A dc current is applied between the ferromagnetic injectors C2 and C5. Since the  contacts are designed lithographically on both sides of the ferromagnetic electrode, the fluctuations in the voltage drop $V_{\text{local}}$ across the flake can be measured via the contact pair C1-C3 (path 1) and C1'-C3' (path 2). The measured signals are cross correlated in order to filter out the noise from external electronics such as preamplifiers and the spectrum analyzer \cite{van_den_brom_quantum_1999}. The electronic 1/f noise $S_{\text{V}}^{\text{local}}$ is measured at different bias currents ($I_{\text{dc}}$) at a fixed carrier density. By fitting the spectrum with the Hooge formula for 1/f noise i.e. $S_{\text{V}}^{\text{local}}=\frac{\gamma^{\text{c}}{ {V}_{\text{local}}}^2}{f^a}$, where $V_{\text{local}}$ is the average voltage drop across the flake and $a$ is the exponent $\sim$~1, we obtain the noise magnitude for the charge transport $\gamma^{\text{c}} \sim 10^{-7}$ (device A in Fig.~\ref{fig: measurement scheme}(c)), similar to the values reported in literature \cite{balandin_low-frequency_2013, pal_microscopic_2011, liu_origin_2013} (see supplementary information for the details). The charge noise magnitude is defined as the Hooge parameter $\gamma_{\text{H}}^{\text{c}}$ divided by the total number of carriers in the transport channel, i.e. $\gamma^{\text{c}}=\gamma_{\text{H}}^{\text{c}}/(n*W*L)$. Here $n$ is charge carrier density, $W$ and $L$ are the width and length of the transport channel. $\gamma^{\text{c}}$ depends both on the concentration and the type of scatterers e.g. short range and long range scatterers \cite{balandin_low-frequency_2013, kaverzin_impurities_2012, pal_microscopic_2011, liu_origin_2013, stolyarov_suppression_2015}.

 The spin-dependent 1/f noise can be expressed as:
 \begin{equation}
  \triangle S_{\text{V}}^{\text{NL}}=\frac{\gamma^{\text{s}} \triangle {V}_{\text{NL}}^2}{f^{a}}=\frac{\gamma^{\text{s}} {({P}{\mu}_{\text{s}}/e)}^2}{f^{a}} 
  \label{eq:Hooge spin}
 \end{equation}
Here $\triangle S_{\text{V}}^{\text{NL}}$ is the spin-dependent non-local noise, $\gamma^{\text{s}}=\gamma_{\text{H}}^{\text{s}}/(n*W*\lambda_{\text{s}})$ is the noise magnitude for spin transport, $e$ is the electronic charge  and $\triangle V_{\text{NL}}= P\mu_{\text{s}}/e$ is the measured non-local spin signal due to the average spin accumulation $\mu_{\text{s}}$ in the channel \footnote{variables $V_{\text{local}}, \triangle V_{\text{NL}}, \mu_{\text{s}}, P, \lambda_{\text{s}}$ represent the time average of the quantities}. Here $\gamma_{\text{H}}^{\text{s}}$ represents the Hooge parameter for spin tranport. In contrast with the charge current, spin current is not a conserved quantity and exists over an effective length scale of $\lambda_{\text{s}}$.  
Spin transport in a non-local geometry is realized in three fundamental steps: i) spin current injection, ii) spin diffusion through the transport channel and iii) detection of the spin accumulation. All these steps can contribute to the spin-dependent noise. For the first step of spin injection,  we use a dc current source to inject spin current, which helps to eliminate the resistance fluctuations in the injector contact, leaving only the polarization fluctuations of the injector electrode as a possible noise source. The polarization fluctuations of the injector can arise due to thermally activated domain wall hopping/rotation in the ferromagnet \cite{jiang_low-frequency_2004,ingvarsson_electronic_1999}. The second possible noise source contributing to the fluctuations in the spin accumulation is the transport channel itself, either via the fluctuating channel resistance or via fluctuations in the spin-relaxation process. The third noise source, similar to the first one, can be present at the detector electrode due to fluctuating contact polarization.   

The spin-dependent noise in graphene is measured non-locally as shown in the connection diagram of Fig.~\ref{fig: measurement scheme}(b). During the noise measurement, we keep the spin injection current I$_{\text{dc}}$ fixed (10 \textmu A) and change the detected spin accumulation in three different  ways. At $B_{\perp}$~=~0~T, i) by changing the spin accumulation by switching the relative magnetization direction of the injector electrodes, ii) by keeping the spin accumulation constant and changing the spin detection sensitivity by switching the relative magnetization direction of detector electrodes, and iii) at $B_{\perp} \neq$~0~T, by dephasing the spins  during transport and thus reducing the spin accumulation. We can also measure the  noise  due to a spin independent background signal at high $B_{\perp} \sim$ 0.12~T, where the spin accumulation is suppressed. The spin-dependent component  $\triangle S_{\text{V}}^{\text{NL}}$ can be estimated by subtracting $S_{\text{V}}^{\text{NL}}$(at $B_{\perp}\sim$ 0.12T) from the measured $S_{\text{V}}^{\text{NL}}$. 

\begin{figure}
\includegraphics[]{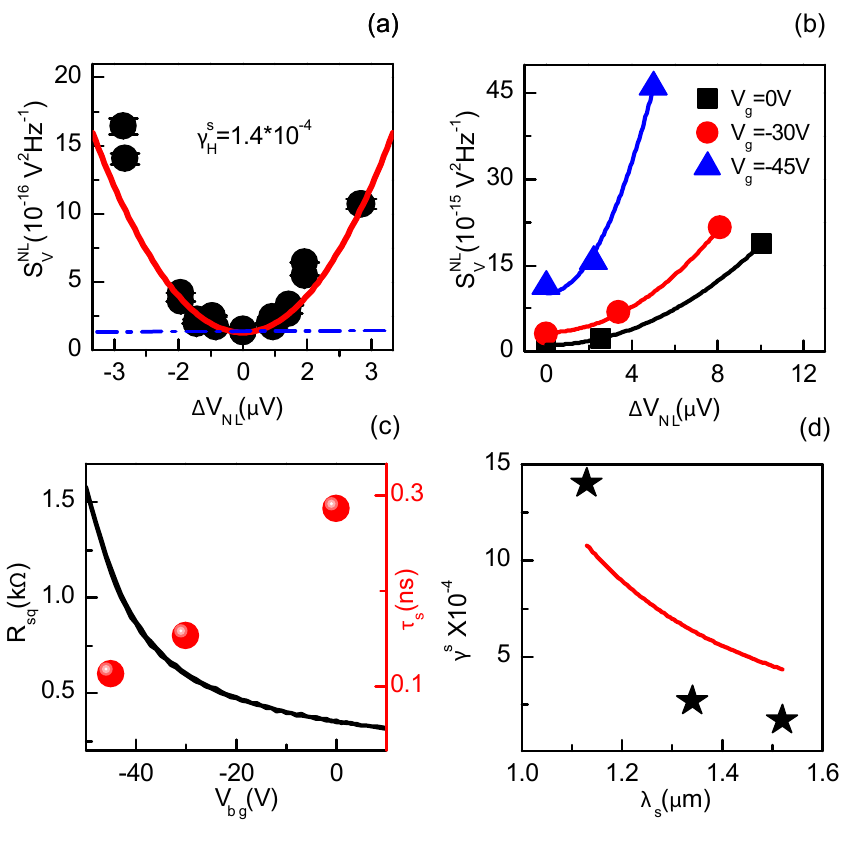}
\caption{\label{fig:spin noise summary} \label{fig:hanle noise gate dependence}(a) summary of noise measured for different spin accumulation potentials in Hanle configuration (device A) and the parabolic fit for the data (red line) using Eq.~\ref{eq:Hooge spin}. Here blue dotted line denotes the spin independent charge noise background i.e.$S_{\text{V}}^{\text{NL}}$(at $B_{\perp} \sim$ 0.12T) (b) Spin dependent 1/f noise in Hanle configuration, measured as a function of back-gate voltage (device B). The increased background noise at $\triangle V_{\text{NL}}$=0 V can come from the charge noise contribution to the non-local signal (S$_{1/f}^{\text{V}} \propto {{I}}^2{R_{\text{sq}}}^2$). (c) The graphene sheet resistance increases at negative back-gate voltage reflecting the n-type doping in graphene and the spin relaxation time (red circles) decreases at lower carrier density for the single layer graphene, resulting in lower value for $\lambda_{\text{s}}$. (d) $\gamma^{\text{s}}$ is increased for lower value of  $\lambda_{\text{s}} (\tau_{\text{s}})$ (black stars), indicating the influence of the spin-flip processes on the extracted noise magnitude for spin transport. A plot of $\gamma^{\text{s}}$ versus $\lambda_{\text{s}}$ with Eq.~\ref{power spectrum analytical} (red curve) shows similar behavior. For the plot, we assume the polarization noise (offset) to be zero, the values for $L$~=1.5~$\mu$m and $S_{\lambda_{\text{s}}} \sim$ 10$^{-16}$ m$^2$Hz$^{-1}$ }
\end{figure}

For the non-local noise measurements in spin valve configuration, the noise PSD measured (Fig.~\ref{fig:spin valve hanle noise}(c)) for the magnetization configuration corresponding to a higher spin accumulation (level II; blue spectrum) is higher in magnitude than for the one corresponding  to a lower spin accumulation (level I; red spectrum) of the spin valve in Fig.~\ref{fig:spin valve hanle noise}(a). In a similar way for the Hanle configuration, we measure the maximum magnitude of the spin-dependent noise for $B_{\perp}$~=~0~T, corresponding to maximum spin accumulation (Fig.~\ref{fig:spin valve hanle noise}(d)). On increasing $|B_{\perp}|$, both the spin accumulation and the associated noise are reduced. 
In order to study its dependence with the spin accumulation, we fit each measured spectrum of $S_V^{\text{NL}}$ versus frequency, obtained  at different spin accumulation values ($\triangle V_{\text{NL}}$) with Eq.~\ref{eq:Hooge spin} in the frequency range of 0.5~Hz-5~Hz. We take the value of $S_{\text{V}}^{\text{NL}}$ at f~=~1~Hz from the fit as a representative value of the 1/f spectrum. The exponent $a$ obtained from the fit is $~ \sim 1$. A summary  of the data points for the noise PSD at different values of spin accumulation, obtained for device A using Hanle precession is plotted in Fig.~\ref{fig:spin noise summary}(a). The $\triangle S_{\text{V}}^{\text{NL}} \propto { {\mu}_{s}}^2$ relation is valid in the lowest order approximation. The parabolic fit of the measured non local noise  using Eq.~\ref{eq:Hooge spin} gives $\gamma^{\text{s}}\sim 10^{-4}$. It should be noted that $\gamma^{\text{s}} \sim 1000\times \gamma^{\text{c}}$, for the same device. Geometrical factors such as length scales cannot account for such a huge difference, as for this sample we obtain $\lambda_{\text{s}} \sim$ 1.5 $\mu$m which is similar to the channel length for charge 1/f noise. The three orders of magnitude enhanced $\gamma^{\text{s}}$ points towards distinctive scattering processes affecting the spin dependent noise, in contrast to the charge 1/f noise. Our findings can be explained along the direction of the recently proposed resonant scattering mechanism \cite{fabian_resonant_scattering} for spin transport where intrinsically present magnetic impurities strongly scatter the spins without a significant effect on the charge scattering strength. The scattering cross section of these impurities can fluctuate in time and could give rise to a spin dependent 1/f noise. 

An analytical expression for the spin-dependent noise (at f~=~1~Hz) which is derived from the equation for the non-local spin signal $\triangle V_{\text{NL}}$ (see supplementary for the complete derivation) can be written as: 
\begin{equation}
  \frac{\triangle S_V^{\text{NL}}}{\triangle V_{\text{NL}}^2} =\gamma^{\text{s}}\simeq \frac{S_P}{{P}^2}+\frac{S_{\lambda_{\text{s}}}}{{\lambda_{\text{s}}}^2}\bigg(1+\frac{L}{{\lambda_{\text{s}}}}\bigg)^2
 \label{power spectrum analytical}
\end{equation}
where $S_P$ is the contact polarization noise which is Fourier transform of the auto correlation function for the time dependent polarization fluctuations i.e. $\mathcal{F}\langle P(t)P(t+\tau)\rangle$ , $S_{\lambda_{\text{s}}}$ is the noise associated with the spin transport i.e. spin relaxation noise ($\mathcal{F}\langle \lambda_{\text{s}}(t)\lambda_{\text{s}}(t+\tau)\rangle$), $L$ is the separation between the inner injector and detector electrodes.
Eq.~\ref{power spectrum analytical} suggests that $\gamma^{\text{s}}$ is increased for lower values of $\lambda_{\text{s}}$. In order to confirm that the spin-dependent noise is affected by the spin transport properties, we measure $S_{\text{V}}^{\text{NL}}$ as a function of the back-gate voltage (carrier density). In agreement with literature \cite{zomer_long-distance_2012, jozsa_linear_2009}, a higher $\tau_{\text{s}}$ is observed at higher charge carrier densities for single layer graphene (Fig.~\ref{fig:hanle noise gate dependence}(c)). The representative data is shown for device B. It is worth emphasizing here that for similar charge and spin transport parameters ($R_{\text{sq}},\lambda_{\text{s}}$) for device A (350~$\Omega$, 1.8~\textmu m) and device B (400~$\Omega$, 1.6~\textmu m), we obtain similar values of $\gamma^{\text{s}}\sim 10^{-4}$. However, both devices have different values of contact polarization $P \sim$ 5\% for device A and $P \sim$ 10\% for device B. This similarity in $\gamma^{\text{s}}$ values despite the difference in $P$ indicates that there is insignificant contribution of the contact polarization noise to the extracted $\gamma^{\text{s}}$. On the other hand, for the noise measurements at different carrier densities, we get an increase in $\gamma^{\text{s}}$ at lower values of $\tau_{\text{s}}$ (Fig.~\ref{fig:hanle noise gate dependence}(d)). The carrier density dependent behavior of the extracted $\gamma^{\text{s}}$ is in qualitative agreement   with the $\lambda_{\text{s}}$ dependence of $\gamma^{\text{s}}$ in Eq.~\ref{power spectrum analytical} (red curve in Fig.~\ref{fig:hanle noise gate dependence}(d)), supporting our hypothesis that the measured spin-dependent noise is dominated by the noise produced by the spin transport (relaxation) process in graphene. 

In order to estimate/filter out the contribution of the contact polarization noise in our measurements, we use spatial cross-correlation (SXC). We measure the contact polarization noise (=$S_{P}/P^2 \times \triangle  {V}_{\text{NL}}^2\sim 10^{-16}$ V$^2$Hz$^{-1}$) which is lower by two orders of magnitude than the spin relaxation noise power between C3 and C5 (= $S_{\lambda_{\text{s}}}/ {\lambda_{\text{s}}}^2\times \triangle V_{\text{NL}}^2\sim 10^{-14}$ V$^2$Hz$^{-1}$)(see supplementary for measurement scheme). Here, based on the reciprocity argument for the injector and detector in spin-valve configuration, we can assume equal noise contribution from the injector electrode and can safely rule out the effect of the polarization noise. 
  

 Since the spin accumulation $ \mu_{\text{s}} \propto \exp^{(-L/\lambda_{\text{s}})}$, the spin relaxation noise is also expected to decay exponentially in accordance with the relation $\triangle S_{\text{V}}^{\text{NL}} \propto  {\mu_{\text{s}}}^2$. We extend our analysis to study the distance dependence of the spin relaxation noise. With the spatial cross-correlation we can also measure the spin relaxation noise between the detector contacts C4 and C5 while removing the polarization noise from contact C4. For this, we measure the spin-dependent noise at different detector contacts via path 1 and path 3 in Fig. \ref{fig: measurement scheme}(b) independently, and cross correlate the measured signals (see supplementary). The polarization noise contribution from the reference detector C5 is expected to be negligible due to the lower value of spin accumulation at the contact ($L_{\text{C1-C5}}/\lambda_{\text{s}}  \sim $ 4).
We measure $\triangle S_{\text{V}}^{\text{NL}}$ at the detectors C3 and C4 for two back-gate voltages: at $V_{\text{g}}$~=~0~V (metallic regime) and at $V_{\text{g}}$~=~-45~V (close to the Dirac point) (see supplementary information).
Using the derived Eq.~\ref{power spectrum analytical}, we can now calculate $\lambda_{\text{s}}$ from  the noise measurement as:
 \begin{equation}
  \frac{S_{\lambda_{\text{s}}}^{\text{C3}}}{S_{\lambda_{\text{s}}}^{\text{C4}}} \simeq {\bigg(\exp{\frac{L^{C3-C4}}{\lambda_{\text{s}}}}\bigg)}^2{\bigg(\frac{1+\frac{L^{\text{C1-C3}}}{\lambda_{\text{s}}}}{1+\frac{L^{\text{C1-C4}}}{\lambda_{\text{s}}}}\bigg)}^2
  \label{eq:noise lambdas}
 \end{equation}
Here $S_{\lambda_{\text{s}}}^{\text{C3}}$ and $S_{\lambda_{\text{s}}}^{\text{C4}}$ are the spin relaxation noise at contacts C3 and C4, and $L^{Ci-Cj}$ is the separation between contacts C$_i$ and C$_j$ (i, j~=~1,3,4). The solution to Eq.~\ref{eq:noise lambdas} for the experimentally obtained noise ratios gives a value of $\lambda_{\text{s}} \sim $1.5~\textmu m and 1.0~\textmu m  at  $V_{\text{g}}$~=~0~V and -45~V, respectively. A close agreement with the values obtained independently from the Hanle measurements ($\lambda_{\text{s}}\sim$~1.5 \textmu m at $V_{\text{g}}$~=~0~V and 1.1~\textmu m at $V_{\text{g}}$~=~-45~V) validates the analytical framework of Eq.~\ref{power spectrum analytical} and Eq.~\ref{eq:noise lambdas}.
 
 By performing the first measurement of 1/f noise associated with spin transport, we demonstrate that the non-local spin-dependent noise in graphene is dominated by the underlying spin relaxation processes. The obtained noise magnitude for charge and spin transport differ by three orders of magnitude, indicating fundamentally different scattering mechanisms such as resonant scattering of the spins, where the fluctuating scattering cross-section of the intrinsically present impurities could produce the spin dependent 1/f fluctuations \cite{fabian_resonant_scattering}. 
The presented work establishes 1/f noise measurements as a complementary approach to extract spin transport parameters and is expected to be valid other spintronic materials, where impurities play an important role in modifying the underlying spin relaxation process.

We acknowledge  J. G. Holstein, H. M. de Roosz and H. Adema for their technical assistance. This research work was financed under EU-graphene flagship program (637088) and supported by the Zernike Institute for Advanced Materials, the Netherlands Organization for Scientific Research (NWO) and the Future and Emerging Technologies (FET) programme within the Seventh Framework Programme  for Research of the European Commission, under FET-open Grant No. 618083 (CN-TQC).


\begin{thebibliography}{33}%
\makeatletter
\providecommand \@ifxundefined [1]{%
 \@ifx{#1\undefined}
}%
\providecommand \@ifnum [1]{%
 \ifnum #1\expandafter \@firstoftwo
 \else \expandafter \@secondoftwo
 \fi
}%
\providecommand \@ifx [1]{%
 \ifx #1\expandafter \@firstoftwo
 \else \expandafter \@secondoftwo
 \fi
}%
\providecommand \natexlab [1]{#1}%
\providecommand \enquote  [1]{``#1''}%
\providecommand \bibnamefont  [1]{#1}%
\providecommand \bibfnamefont [1]{#1}%
\providecommand \citenamefont [1]{#1}%
\providecommand \href@noop [0]{\@secondoftwo}%
\providecommand \href [0]{\begingroup \@sanitize@url \@href}%
\providecommand \@href[1]{\@@startlink{#1}\@@href}%
\providecommand \@@href[1]{\endgroup#1\@@endlink}%
\providecommand \@sanitize@url [0]{\catcode `\\12\catcode `\$12\catcode
  `\&12\catcode `\#12\catcode `\^12\catcode `\_12\catcode `\%12\relax}%
\providecommand \@@startlink[1]{}%
\providecommand \@@endlink[0]{}%
\providecommand \url  [0]{\begingroup\@sanitize@url \@url }%
\providecommand \@url [1]{\endgroup\@href {#1}{\urlprefix }}%
\providecommand \urlprefix  [0]{URL }%
\providecommand \Eprint [0]{\href }%
\providecommand \doibase [0]{http://dx.doi.org/}%
\providecommand \selectlanguage [0]{\@gobble}%
\providecommand \bibinfo  [0]{\@secondoftwo}%
\providecommand \bibfield  [0]{\@secondoftwo}%
\providecommand \translation [1]{[#1]}%
\providecommand \BibitemOpen [0]{}%
\providecommand \bibitemStop [0]{}%
\providecommand \bibitemNoStop [0]{.\EOS\space}%
\providecommand \EOS [0]{\spacefactor3000\relax}%
\providecommand \BibitemShut  [1]{\csname bibitem#1\endcsname}%
\let\auto@bib@innerbib\@empty
\bibitem [{\citenamefont {Landauer}(1998)}]{landauer_condensed-matter_1998}%
  \BibitemOpen
  \bibfield  {author} {\bibinfo {author} {\bibfnamefont {R.}~\bibnamefont
  {Landauer}},\ }\href {\doibase 10.1038/33551} {\bibfield  {journal} {\bibinfo
   {journal} {Nature}\ }\textbf {\bibinfo {volume} {392}},\ \bibinfo {pages}
  {658} (\bibinfo {year} {1998})}\BibitemShut {NoStop}%
\bibitem [{\citenamefont {Jayaraman}\ and\ \citenamefont
  {Sodini}(1989)}]{jayaraman_1/f_1989}%
  \BibitemOpen
  \bibfield  {author} {\bibinfo {author} {\bibfnamefont {R.}~\bibnamefont
  {Jayaraman}}\ and\ \bibinfo {author} {\bibfnamefont {C.}~\bibnamefont
  {Sodini}},\ }\href {\doibase 10.1109/16.34242} {\bibfield  {journal}
  {\bibinfo  {journal} {IEEE Trans. Electron Devices}\ }\textbf {\bibinfo
  {volume} {36}},\ \bibinfo {pages} {1773} (\bibinfo {year}
  {1989})}\BibitemShut {NoStop}%
\bibitem [{\citenamefont {Hooge}\ \emph {et~al.}(1981)\citenamefont {Hooge},
  \citenamefont {Kleinpenning},\ and\ \citenamefont
  {Vandamme}}]{hooge_experimental_1981}%
  \BibitemOpen
  \bibfield  {author} {\bibinfo {author} {\bibfnamefont {F.~N.}\ \bibnamefont
  {Hooge}}, \bibinfo {author} {\bibfnamefont {T.~G.~M.}\ \bibnamefont
  {Kleinpenning}}, \ and\ \bibinfo {author} {\bibfnamefont {L.~K.~J.}\
  \bibnamefont {Vandamme}},\ }\href {\doibase 10.1088/0034-4885/44/5/001}
  {\bibfield  {journal} {\bibinfo  {journal} {Rep. Prog. Phys.}\ }\textbf
  {\bibinfo {volume} {44}},\ \bibinfo {pages} {479} (\bibinfo {year}
  {1981})}\BibitemShut {NoStop}%
\bibitem [{\citenamefont {Dutta}\ and\ \citenamefont
  {Horn}(1981)}]{dutta_low-frequency_1981}%
  \BibitemOpen
  \bibfield  {author} {\bibinfo {author} {\bibfnamefont {P.}~\bibnamefont
  {Dutta}}\ and\ \bibinfo {author} {\bibfnamefont {P.~M.}\ \bibnamefont
  {Horn}},\ }\href {\doibase 10.1103/RevModPhys.53.497} {\bibfield  {journal}
  {\bibinfo  {journal} {Rev. Mod. Phys.}\ }\textbf {\bibinfo {volume} {53}},\
  \bibinfo {pages} {497} (\bibinfo {year} {1981})}\BibitemShut {NoStop}%
\bibitem [{\citenamefont {Ertler}\ \emph
  {et~al.}(2009{\natexlab{a}})\citenamefont {Ertler}, \citenamefont {Konschuh},
  \citenamefont {Gmitra},\ and\ \citenamefont {Fabian}}]{ertler_electron_2009}%
  \BibitemOpen
  \bibfield  {author} {\bibinfo {author} {\bibfnamefont {C.}~\bibnamefont
  {Ertler}}, \bibinfo {author} {\bibfnamefont {S.}~\bibnamefont {Konschuh}},
  \bibinfo {author} {\bibfnamefont {M.}~\bibnamefont {Gmitra}}, \ and\ \bibinfo
  {author} {\bibfnamefont {J.}~\bibnamefont {Fabian}},\ }\href {\doibase
  10.1103/PhysRevB.80.041405} {\bibfield  {journal} {\bibinfo  {journal} {Phys.
  Rev. B}\ }\textbf {\bibinfo {volume} {80}},\ \bibinfo {pages} {041405}
  (\bibinfo {year} {2009}{\natexlab{a}})}\BibitemShut {NoStop}%
\bibitem [{\citenamefont {Huertas-Hernando}\ \emph {et~al.}(2009)\citenamefont
  {Huertas-Hernando}, \citenamefont {Guinea},\ and\ \citenamefont
  {Brataas}}]{huertas-hernando_spin-orbit-mediated_2009}%
  \BibitemOpen
  \bibfield  {author} {\bibinfo {author} {\bibfnamefont {D.}~\bibnamefont
  {Huertas-Hernando}}, \bibinfo {author} {\bibfnamefont {F.}~\bibnamefont
  {Guinea}}, \ and\ \bibinfo {author} {\bibfnamefont {A.}~\bibnamefont
  {Brataas}},\ }\href {\doibase 10.1103/PhysRevLett.103.146801} {\bibfield
  {journal} {\bibinfo  {journal} {Phys. Rev. Lett.}\ }\textbf {\bibinfo
  {volume} {103}},\ \bibinfo {pages} {146801} (\bibinfo {year}
  {2009})}\BibitemShut {NoStop}%
\bibitem [{\citenamefont {Ingla-Ayn\'{e}s}\ \emph {et~al.}(2015)\citenamefont
  {Ingla-Ayn\'{e}s}, \citenamefont {Guimar\~{a}es}, \citenamefont {Meijerink},
  \citenamefont {Zomer},\ and\ \citenamefont {van Wees}}]{pep_2015}%
  \BibitemOpen
  \bibfield  {author} {\bibinfo {author} {\bibfnamefont {J.}~\bibnamefont
  {Ingla-Ayn\'{e}s}}, \bibinfo {author} {\bibfnamefont {M.~H.~D.}\ \bibnamefont
  {Guimar\~{a}es}}, \bibinfo {author} {\bibfnamefont {R.~J.}\ \bibnamefont
  {Meijerink}}, \bibinfo {author} {\bibfnamefont {P.~J.}\ \bibnamefont
  {Zomer}}, \ and\ \bibinfo {author} {\bibfnamefont {B.~J.}\ \bibnamefont {van
  Wees}},\ }\href {\doibase 10.1103/PhysRevB.92.201410} {\bibfield  {journal}
  {\bibinfo  {journal} {Phys. Rev. B}\ }\textbf {\bibinfo {volume} {92}},\
  \bibinfo {pages} {201410} (\bibinfo {year} {2015})}\BibitemShut {NoStop}%
\bibitem [{\citenamefont {Dugaev}\ \emph {et~al.}(2011)\citenamefont {Dugaev},
  \citenamefont {Sherman},\ and\ \citenamefont {Barna\'{s}}}]{branas_graphene}%
  \BibitemOpen
  \bibfield  {author} {\bibinfo {author} {\bibfnamefont {V.~K.}\ \bibnamefont
  {Dugaev}}, \bibinfo {author} {\bibfnamefont {E.~Y.}\ \bibnamefont {Sherman}},
  \ and\ \bibinfo {author} {\bibfnamefont {J.}~\bibnamefont {Barna\'{s}}},\
  }\href {\doibase 10.1103/PhysRevB.83.085306} {\bibfield  {journal} {\bibinfo
  {journal} {Phys. Rev. B}\ }\textbf {\bibinfo {volume} {83}},\ \bibinfo
  {pages} {085306} (\bibinfo {year} {2011})}\BibitemShut {NoStop}%
\bibitem [{\citenamefont {Min}\ \emph {et~al.}(2006)\citenamefont {Min},
  \citenamefont {Hill}, \citenamefont {Sinitsyn}, \citenamefont {Sahu},
  \citenamefont {Kleinman},\ and\ \citenamefont
  {MacDonald}}]{macdonald_graphene}%
  \BibitemOpen
  \bibfield  {author} {\bibinfo {author} {\bibfnamefont {H.}~\bibnamefont
  {Min}}, \bibinfo {author} {\bibfnamefont {J.~E.}\ \bibnamefont {Hill}},
  \bibinfo {author} {\bibfnamefont {N.~A.}\ \bibnamefont {Sinitsyn}}, \bibinfo
  {author} {\bibfnamefont {B.~R.}\ \bibnamefont {Sahu}}, \bibinfo {author}
  {\bibfnamefont {L.}~\bibnamefont {Kleinman}}, \ and\ \bibinfo {author}
  {\bibfnamefont {A.~H.}\ \bibnamefont {MacDonald}},\ }\href {\doibase
  10.1103/PhysRevB.74.165310} {\bibfield  {journal} {\bibinfo  {journal} {Phys.
  Rev. B}\ }\textbf {\bibinfo {volume} {74}},\ \bibinfo {pages} {165310}
  (\bibinfo {year} {2006})}\BibitemShut {NoStop}%
\bibitem [{\citenamefont {Ertler}\ \emph
  {et~al.}(2009{\natexlab{b}})\citenamefont {Ertler}, \citenamefont {Konschuh},
  \citenamefont {Gmitra},\ and\ \citenamefont
  {Fabian}}]{fabian_relaxation_sub}%
  \BibitemOpen
  \bibfield  {author} {\bibinfo {author} {\bibfnamefont {C.}~\bibnamefont
  {Ertler}}, \bibinfo {author} {\bibfnamefont {S.}~\bibnamefont {Konschuh}},
  \bibinfo {author} {\bibfnamefont {M.}~\bibnamefont {Gmitra}}, \ and\ \bibinfo
  {author} {\bibfnamefont {J.}~\bibnamefont {Fabian}},\ }\href {\doibase
  10.1103/PhysRevB.80.041405} {\bibfield  {journal} {\bibinfo  {journal} {Phys.
  Rev. B}\ }\textbf {\bibinfo {volume} {80}},\ \bibinfo {pages} {041405}
  (\bibinfo {year} {2009}{\natexlab{b}})}\BibitemShut {NoStop}%
\bibitem [{\citenamefont {Lundeberg}\ \emph {et~al.}(2013)\citenamefont
  {Lundeberg}, \citenamefont {Yang}, \citenamefont {Renard},\ and\
  \citenamefont {Folk}}]{Folk_relaxation}%
  \BibitemOpen
  \bibfield  {author} {\bibinfo {author} {\bibfnamefont {M.~B.}\ \bibnamefont
  {Lundeberg}}, \bibinfo {author} {\bibfnamefont {R.}~\bibnamefont {Yang}},
  \bibinfo {author} {\bibfnamefont {J.}~\bibnamefont {Renard}}, \ and\ \bibinfo
  {author} {\bibfnamefont {J.~A.}\ \bibnamefont {Folk}},\ }\href {\doibase
  10.1103/PhysRevLett.110.156601} {\bibfield  {journal} {\bibinfo  {journal}
  {Phys. Rev. Lett.}\ }\textbf {\bibinfo {volume} {110}},\ \bibinfo {pages}
  {156601} (\bibinfo {year} {2013})}\BibitemShut {NoStop}%
\bibitem [{\citenamefont {Kochan}\ \emph {et~al.}(2014)\citenamefont {Kochan},
  \citenamefont {Gmitra},\ and\ \citenamefont
  {Fabian}}]{fabian_resonant_scattering}%
  \BibitemOpen
  \bibfield  {author} {\bibinfo {author} {\bibfnamefont {D.}~\bibnamefont
  {Kochan}}, \bibinfo {author} {\bibfnamefont {M.}~\bibnamefont {Gmitra}}, \
  and\ \bibinfo {author} {\bibfnamefont {J.}~\bibnamefont {Fabian}},\ }\href
  {\doibase 10.1103/PhysRevLett.112.116602} {\bibfield  {journal} {\bibinfo
  {journal} {Phys. Rev. Lett.}\ }\textbf {\bibinfo {volume} {112}},\ \bibinfo
  {pages} {116602} (\bibinfo {year} {2014})}\BibitemShut {NoStop}%
\bibitem [{\citenamefont {Soriano}\ \emph {et~al.}(2015)\citenamefont
  {Soriano}, \citenamefont {Tuan}, \citenamefont {Dubois}, \citenamefont
  {Gmitra}, \citenamefont {Cummings}, \citenamefont {Kochan}, \citenamefont
  {Ortmann}, \citenamefont {Charlier}, \citenamefont {Fabian},\ and\
  \citenamefont {Roche}}]{review_Roche}%
  \BibitemOpen
  \bibfield  {author} {\bibinfo {author} {\bibfnamefont {D.}~\bibnamefont
  {Soriano}}, \bibinfo {author} {\bibfnamefont {D.~V.}\ \bibnamefont {Tuan}},
  \bibinfo {author} {\bibfnamefont {S.~M.-M.}\ \bibnamefont {Dubois}}, \bibinfo
  {author} {\bibfnamefont {M.}~\bibnamefont {Gmitra}}, \bibinfo {author}
  {\bibfnamefont {A.~W.}\ \bibnamefont {Cummings}}, \bibinfo {author}
  {\bibfnamefont {D.}~\bibnamefont {Kochan}}, \bibinfo {author} {\bibfnamefont
  {F.}~\bibnamefont {Ortmann}}, \bibinfo {author} {\bibfnamefont {J.-C.}\
  \bibnamefont {Charlier}}, \bibinfo {author} {\bibfnamefont {J.}~\bibnamefont
  {Fabian}}, \ and\ \bibinfo {author} {\bibfnamefont {S.}~\bibnamefont
  {Roche}},\ }\href {\doibase 10.1088/2053-1583/2/2/022002} {\bibfield
  {journal} {\bibinfo  {journal} {2D Mater.}\ }\textbf {\bibinfo {volume}
  {2}},\ \bibinfo {pages} {022002} (\bibinfo {year} {2015})}\BibitemShut
  {NoStop}%
\bibitem [{\citenamefont {Omar}\ \emph {et~al.}(2015)\citenamefont {Omar},
  \citenamefont {Gurram}, \citenamefont {Vera-Marun}, \citenamefont {Zhang},
  \citenamefont {Huisman}, \citenamefont {Kaverzin}, \citenamefont {Feringa},\
  and\ \citenamefont {van Wees}}]{omar_spin_2015}%
  \BibitemOpen
  \bibfield  {author} {\bibinfo {author} {\bibfnamefont {S.}~\bibnamefont
  {Omar}}, \bibinfo {author} {\bibfnamefont {M.}~\bibnamefont {Gurram}},
  \bibinfo {author} {\bibfnamefont {I.~J.}\ \bibnamefont {Vera-Marun}},
  \bibinfo {author} {\bibfnamefont {X.}~\bibnamefont {Zhang}}, \bibinfo
  {author} {\bibfnamefont {E.~H.}\ \bibnamefont {Huisman}}, \bibinfo {author}
  {\bibfnamefont {A.}~\bibnamefont {Kaverzin}}, \bibinfo {author}
  {\bibfnamefont {B.~L.}\ \bibnamefont {Feringa}}, \ and\ \bibinfo {author}
  {\bibfnamefont {B.~J.}\ \bibnamefont {van Wees}},\ }\href {\doibase
  10.1103/PhysRevB.92.115442} {\bibfield  {journal} {\bibinfo  {journal} {Phys.
  Rev. B}\ }\textbf {\bibinfo {volume} {92}},\ \bibinfo {pages} {115442}
  (\bibinfo {year} {2015})}\BibitemShut {NoStop}%
\bibitem [{\citenamefont {Arakawa}\ \emph {et~al.}(2015)\citenamefont
  {Arakawa}, \citenamefont {Shiogai}, \citenamefont {Ciorga}, \citenamefont
  {Utz}, \citenamefont {Schuh}, \citenamefont {Kohda}, \citenamefont {Nitta},
  \citenamefont {Bougeard}, \citenamefont {Weiss}, \citenamefont {Ono},\ and\
  \citenamefont {Kobayashi}}]{arakawa_shot_2015}%
  \BibitemOpen
  \bibfield  {author} {\bibinfo {author} {\bibfnamefont {T.}~\bibnamefont
  {Arakawa}}, \bibinfo {author} {\bibfnamefont {J.}~\bibnamefont {Shiogai}},
  \bibinfo {author} {\bibfnamefont {M.}~\bibnamefont {Ciorga}}, \bibinfo
  {author} {\bibfnamefont {M.}~\bibnamefont {Utz}}, \bibinfo {author}
  {\bibfnamefont {D.}~\bibnamefont {Schuh}}, \bibinfo {author} {\bibfnamefont
  {M.}~\bibnamefont {Kohda}}, \bibinfo {author} {\bibfnamefont
  {J.}~\bibnamefont {Nitta}}, \bibinfo {author} {\bibfnamefont
  {D.}~\bibnamefont {Bougeard}}, \bibinfo {author} {\bibfnamefont
  {D.}~\bibnamefont {Weiss}}, \bibinfo {author} {\bibfnamefont
  {T.}~\bibnamefont {Ono}}, \ and\ \bibinfo {author} {\bibfnamefont
  {K.}~\bibnamefont {Kobayashi}},\ }\href {\doibase
  10.1103/PhysRevLett.114.016601} {\bibfield  {journal} {\bibinfo  {journal}
  {Phys. Rev. Lett.}\ }\textbf {\bibinfo {volume} {114}},\ \bibinfo {pages}
  {016601} (\bibinfo {year} {2015})}\BibitemShut {NoStop}%
\bibitem [{\citenamefont {Tombros}\ \emph {et~al.}(2007)\citenamefont
  {Tombros}, \citenamefont {J\~{o}zsa}, \citenamefont {Popinciuc},
  \citenamefont {Jonkman},\ and\ \citenamefont {van Wees}}]{Tombros_nature}%
  \BibitemOpen
  \bibfield  {author} {\bibinfo {author} {\bibfnamefont {N.}~\bibnamefont
  {Tombros}}, \bibinfo {author} {\bibfnamefont {C.}~\bibnamefont {J\~{o}zsa}},
  \bibinfo {author} {\bibfnamefont {M.}~\bibnamefont {Popinciuc}}, \bibinfo
  {author} {\bibfnamefont {H.~T.}\ \bibnamefont {Jonkman}}, \ and\ \bibinfo
  {author} {\bibfnamefont {B.~J.}\ \bibnamefont {van Wees}},\ }\href {\doibase
  10.1038/nature06037} {\bibfield  {journal} {\bibinfo  {journal} {Nature}\
  }\textbf {\bibinfo {volume} {448}},\ \bibinfo {pages} {571} (\bibinfo {year}
  {2007})}\BibitemShut {NoStop}%
\bibitem [{\citenamefont {van~den Brom}\ and\ \citenamefont {van
  Ruitenbeek}(1999)}]{van_den_brom_quantum_1999}%
  \BibitemOpen
  \bibfield  {author} {\bibinfo {author} {\bibfnamefont {H.~E.}\ \bibnamefont
  {van~den Brom}}\ and\ \bibinfo {author} {\bibfnamefont {J.~M.}\ \bibnamefont
  {van Ruitenbeek}},\ }\href {\doibase 10.1103/PhysRevLett.82.1526} {\bibfield
  {journal} {\bibinfo  {journal} {Phys. Rev. Lett.}\ }\textbf {\bibinfo
  {volume} {82}},\ \bibinfo {pages} {1526} (\bibinfo {year}
  {1999})}\BibitemShut {NoStop}%
\bibitem [{\citenamefont {Balandin}(2013)}]{balandin_low-frequency_2013}%
  \BibitemOpen
  \bibfield  {author} {\bibinfo {author} {\bibfnamefont {A.~A.}\ \bibnamefont
  {Balandin}},\ }\href {\doibase 10.1038/nnano.2013.144} {\bibfield  {journal}
  {\bibinfo  {journal} {Nat Nano}\ }\textbf {\bibinfo {volume} {8}},\ \bibinfo
  {pages} {549} (\bibinfo {year} {2013})}\BibitemShut {NoStop}%
\bibitem [{\citenamefont {Pal}\ \emph {et~al.}(2011)\citenamefont {Pal},
  \citenamefont {Ghatak}, \citenamefont {Kochat}, \citenamefont {Sneha},
  \citenamefont {Sampathkumar}, \citenamefont {Raghavan},\ and\ \citenamefont
  {Ghosh}}]{pal_microscopic_2011}%
  \BibitemOpen
  \bibfield  {author} {\bibinfo {author} {\bibfnamefont {A.~N.}\ \bibnamefont
  {Pal}}, \bibinfo {author} {\bibfnamefont {S.}~\bibnamefont {Ghatak}},
  \bibinfo {author} {\bibfnamefont {V.}~\bibnamefont {Kochat}}, \bibinfo
  {author} {\bibfnamefont {E.~S.}\ \bibnamefont {Sneha}}, \bibinfo {author}
  {\bibfnamefont {A.}~\bibnamefont {Sampathkumar}}, \bibinfo {author}
  {\bibfnamefont {S.}~\bibnamefont {Raghavan}}, \ and\ \bibinfo {author}
  {\bibfnamefont {A.}~\bibnamefont {Ghosh}},\ }\href {\doibase
  10.1021/nn103273n} {\bibfield  {journal} {\bibinfo  {journal} {ACS Nano}\
  }\textbf {\bibinfo {volume} {5}},\ \bibinfo {pages} {2075} (\bibinfo {year}
  {2011})}\BibitemShut {NoStop}%
\bibitem [{\citenamefont {Liu}\ \emph {et~al.}(2013)\citenamefont {Liu},
  \citenamefont {Rumyantsev}, \citenamefont {Shur},\ and\ \citenamefont
  {Balandin}}]{liu_origin_2013}%
  \BibitemOpen
  \bibfield  {author} {\bibinfo {author} {\bibfnamefont {G.}~\bibnamefont
  {Liu}}, \bibinfo {author} {\bibfnamefont {S.}~\bibnamefont {Rumyantsev}},
  \bibinfo {author} {\bibfnamefont {M.~S.}\ \bibnamefont {Shur}}, \ and\
  \bibinfo {author} {\bibfnamefont {A.~A.}\ \bibnamefont {Balandin}},\ }\href
  {\doibase 10.1063/1.4794843} {\bibfield  {journal} {\bibinfo  {journal}
  {Appl. Phys. Lett.}\ }\textbf {\bibinfo {volume} {102}},\ \bibinfo {pages}
  {093111} (\bibinfo {year} {2013})}\BibitemShut {NoStop}%
\bibitem [{\citenamefont {Kaverzin}\ \emph {et~al.}(2012)\citenamefont
  {Kaverzin}, \citenamefont {Mayorov}, \citenamefont {Shytov},\ and\
  \citenamefont {Horsell}}]{kaverzin_impurities_2012}%
  \BibitemOpen
  \bibfield  {author} {\bibinfo {author} {\bibfnamefont {A.~A.}\ \bibnamefont
  {Kaverzin}}, \bibinfo {author} {\bibfnamefont {A.~S.}\ \bibnamefont
  {Mayorov}}, \bibinfo {author} {\bibfnamefont {A.}~\bibnamefont {Shytov}}, \
  and\ \bibinfo {author} {\bibfnamefont {D.~W.}\ \bibnamefont {Horsell}},\
  }\href {\doibase 10.1103/PhysRevB.85.075435} {\bibfield  {journal} {\bibinfo
  {journal} {Phys. Rev. B}\ }\textbf {\bibinfo {volume} {85}},\ \bibinfo
  {pages} {075435} (\bibinfo {year} {2012})}\BibitemShut {NoStop}%
\bibitem [{\citenamefont {Stolyarov}\ \emph {et~al.}(2015)\citenamefont
  {Stolyarov}, \citenamefont {Liu}, \citenamefont {Rumyantsev}, \citenamefont
  {Shur},\ and\ \citenamefont {Balandin}}]{stolyarov_suppression_2015}%
  \BibitemOpen
  \bibfield  {author} {\bibinfo {author} {\bibfnamefont {M.~A.}\ \bibnamefont
  {Stolyarov}}, \bibinfo {author} {\bibfnamefont {G.}~\bibnamefont {Liu}},
  \bibinfo {author} {\bibfnamefont {S.~L.}\ \bibnamefont {Rumyantsev}},
  \bibinfo {author} {\bibfnamefont {M.}~\bibnamefont {Shur}}, \ and\ \bibinfo
  {author} {\bibfnamefont {A.~A.}\ \bibnamefont {Balandin}},\ }\href {\doibase
  10.1063/1.4926872} {\bibfield  {journal} {\bibinfo  {journal} {Appl. Phys.
  Lett.}\ }\textbf {\bibinfo {volume} {107}},\ \bibinfo {pages} {023106}
  (\bibinfo {year} {2015})}\BibitemShut {NoStop}%
\bibitem [{Note1()}]{Note1}%
  \BibitemOpen
  \bibinfo {note} {Variables $V_{\protect \text {local}}, \triangle V_{\protect
  \text {NL}}, \mu _{\protect \text {s}}, P, \lambda _{\protect \text {s}}$
  represent the time average of the quantities}\BibitemShut {NoStop}%
\bibitem [{\citenamefont {Jiang}\ \emph {et~al.}(2004)\citenamefont {Jiang},
  \citenamefont {Nowak}, \citenamefont {Scott}, \citenamefont {Johnson},
  \citenamefont {Slaughter}, \citenamefont {Sun},\ and\ \citenamefont
  {Dave}}]{jiang_low-frequency_2004}%
  \BibitemOpen
  \bibfield  {author} {\bibinfo {author} {\bibfnamefont {L.}~\bibnamefont
  {Jiang}}, \bibinfo {author} {\bibfnamefont {E.~R.}\ \bibnamefont {Nowak}},
  \bibinfo {author} {\bibfnamefont {P.~E.}\ \bibnamefont {Scott}}, \bibinfo
  {author} {\bibfnamefont {J.}~\bibnamefont {Johnson}}, \bibinfo {author}
  {\bibfnamefont {J.~M.}\ \bibnamefont {Slaughter}}, \bibinfo {author}
  {\bibfnamefont {J.~J.}\ \bibnamefont {Sun}}, \ and\ \bibinfo {author}
  {\bibfnamefont {R.~W.}\ \bibnamefont {Dave}},\ }\href {\doibase
  10.1103/PhysRevB.69.054407} {\bibfield  {journal} {\bibinfo  {journal} {Phys.
  Rev. B}\ }\textbf {\bibinfo {volume} {69}},\ \bibinfo {pages} {054407}
  (\bibinfo {year} {2004})}\BibitemShut {NoStop}%
\bibitem [{\citenamefont {Ingvarsson}\ \emph {et~al.}(1999)\citenamefont
  {Ingvarsson}, \citenamefont {Xiao}, \citenamefont {Wanner}, \citenamefont
  {Trouilloud}, \citenamefont {Lu}, \citenamefont {Gallagher}, \citenamefont
  {Marley}, \citenamefont {Roche},\ and\ \citenamefont
  {Parkin}}]{ingvarsson_electronic_1999}%
  \BibitemOpen
  \bibfield  {author} {\bibinfo {author} {\bibfnamefont {S.}~\bibnamefont
  {Ingvarsson}}, \bibinfo {author} {\bibfnamefont {G.}~\bibnamefont {Xiao}},
  \bibinfo {author} {\bibfnamefont {R.~A.}\ \bibnamefont {Wanner}}, \bibinfo
  {author} {\bibfnamefont {P.}~\bibnamefont {Trouilloud}}, \bibinfo {author}
  {\bibfnamefont {Y.}~\bibnamefont {Lu}}, \bibinfo {author} {\bibfnamefont
  {W.~J.}\ \bibnamefont {Gallagher}}, \bibinfo {author} {\bibfnamefont
  {A.}~\bibnamefont {Marley}}, \bibinfo {author} {\bibfnamefont {K.~P.}\
  \bibnamefont {Roche}}, \ and\ \bibinfo {author} {\bibfnamefont {S.~S.~P.}\
  \bibnamefont {Parkin}},\ }\href {\doibase 10.1063/1.369851} {\bibfield
  {journal} {\bibinfo  {journal} {J. Appl. Phys.}\ }\textbf {\bibinfo {volume}
  {85}},\ \bibinfo {pages} {5270} (\bibinfo {year} {1999})}\BibitemShut
  {NoStop}%
\bibitem [{\citenamefont {Zomer}\ \emph {et~al.}(2012)\citenamefont {Zomer},
  \citenamefont {Guimar\~{a}es}, \citenamefont {Tombros},\ and\ \citenamefont
  {van Wees}}]{zomer_long-distance_2012}%
  \BibitemOpen
  \bibfield  {author} {\bibinfo {author} {\bibfnamefont {P.~J.}\ \bibnamefont
  {Zomer}}, \bibinfo {author} {\bibfnamefont {M.~H.~D.}\ \bibnamefont
  {Guimar\~{a}es}}, \bibinfo {author} {\bibfnamefont {N.}~\bibnamefont
  {Tombros}}, \ and\ \bibinfo {author} {\bibfnamefont {B.~J.}\ \bibnamefont
  {van Wees}},\ }\href {\doibase 10.1103/PhysRevB.86.161416} {\bibfield
  {journal} {\bibinfo  {journal} {Phys. Rev. B}\ }\textbf {\bibinfo {volume}
  {86}},\ \bibinfo {pages} {161416} (\bibinfo {year} {2012})}\BibitemShut
  {NoStop}%
\bibitem [{\citenamefont {J\~{o}zsa}\ \emph {et~al.}(2009)\citenamefont
  {J\~{o}zsa}, \citenamefont {Maassen}, \citenamefont {Popinciuc},
  \citenamefont {Zomer}, \citenamefont {Veligura}, \citenamefont {Jonkman},\
  and\ \citenamefont {van Wees}}]{jozsa_linear_2009}%
  \BibitemOpen
  \bibfield  {author} {\bibinfo {author} {\bibfnamefont {C.}~\bibnamefont
  {J\~{o}zsa}}, \bibinfo {author} {\bibfnamefont {T.}~\bibnamefont {Maassen}},
  \bibinfo {author} {\bibfnamefont {M.}~\bibnamefont {Popinciuc}}, \bibinfo
  {author} {\bibfnamefont {P.~J.}\ \bibnamefont {Zomer}}, \bibinfo {author}
  {\bibfnamefont {A.}~\bibnamefont {Veligura}}, \bibinfo {author}
  {\bibfnamefont {H.~T.}\ \bibnamefont {Jonkman}}, \ and\ \bibinfo {author}
  {\bibfnamefont {B.~J.}\ \bibnamefont {van Wees}},\ }\href {\doibase
  10.1103/PhysRevB.80.241403} {\bibfield  {journal} {\bibinfo  {journal} {Phys.
  Rev. B}\ }\textbf {\bibinfo {volume} {80}},\ \bibinfo {pages} {241403}
  (\bibinfo {year} {2009})}\BibitemShut {NoStop}%
\bibitem [{\citenamefont {Maassen}\ \emph {et~al.}(2012)\citenamefont
  {Maassen}, \citenamefont {Vera-Marun}, \citenamefont {Guimar\~{a}es},\ and\
  \citenamefont {van Wees}}]{maassen_contacts}%
  \BibitemOpen
  \bibfield  {author} {\bibinfo {author} {\bibfnamefont {T.}~\bibnamefont
  {Maassen}}, \bibinfo {author} {\bibfnamefont {I.~J.}\ \bibnamefont
  {Vera-Marun}}, \bibinfo {author} {\bibfnamefont {M.~H.~D.}\ \bibnamefont
  {Guimar\~{a}es}}, \ and\ \bibinfo {author} {\bibfnamefont {B.~J.}\
  \bibnamefont {van Wees}},\ }\href {\doibase 10.1103/PhysRevB.86.235408}
  {\bibfield  {journal} {\bibinfo  {journal} {Phys. Rev. B}\ }\textbf {\bibinfo
  {volume} {86}},\ \bibinfo {pages} {235408} (\bibinfo {year}
  {2012})}\BibitemShut {NoStop}%
\bibitem [{\citenamefont {Hooge}(1994)}]{hooge_1/f_1994}%
  \BibitemOpen
  \bibfield  {author} {\bibinfo {author} {\bibfnamefont {F.}~\bibnamefont
  {Hooge}},\ }\href {\doibase 10.1109/16.333808} {\bibfield  {journal}
  {\bibinfo  {journal} {IEEE Trans. Electron Devices}\ }\textbf {\bibinfo
  {volume} {41}},\ \bibinfo {pages} {1926} (\bibinfo {year}
  {1994})}\BibitemShut {NoStop}%
\bibitem [{\citenamefont {Berger}\ \emph {et~al.}(2015)\citenamefont {Berger},
  \citenamefont {Page}, \citenamefont {Wen}, \citenamefont {McCreary},
  \citenamefont {Bhallamudi}, \citenamefont {Kawakami},\ and\ \citenamefont
  {Chris~Hammel}}]{berger_correlating_2015}%
  \BibitemOpen
  \bibfield  {author} {\bibinfo {author} {\bibfnamefont {A.~J.}\ \bibnamefont
  {Berger}}, \bibinfo {author} {\bibfnamefont {M.~R.}\ \bibnamefont {Page}},
  \bibinfo {author} {\bibfnamefont {H.}~\bibnamefont {Wen}}, \bibinfo {author}
  {\bibfnamefont {K.~M.}\ \bibnamefont {McCreary}}, \bibinfo {author}
  {\bibfnamefont {V.~P.}\ \bibnamefont {Bhallamudi}}, \bibinfo {author}
  {\bibfnamefont {R.~K.}\ \bibnamefont {Kawakami}}, \ and\ \bibinfo {author}
  {\bibfnamefont {P.}~\bibnamefont {Chris~Hammel}},\ }\href {\doibase
  10.1063/1.4932673} {\bibfield  {journal} {\bibinfo  {journal} {Appl. Phys.
  Lett.}\ }\textbf {\bibinfo {volume} {107}},\ \bibinfo {pages} {142406}
  (\bibinfo {year} {2015})}\BibitemShut {NoStop}%
\bibitem [{\citenamefont {Guimarães}\ \emph {et~al.}(2014)\citenamefont
  {Guimarães}, \citenamefont {van~den Berg}, \citenamefont {Vera-Marun},
  \citenamefont {Zomer},\ and\ \citenamefont {van Wees}}]{guimaraes_spin_2014}%
  \BibitemOpen
  \bibfield  {author} {\bibinfo {author} {\bibfnamefont {M.~H.~D.}\
  \bibnamefont {Guimarães}}, \bibinfo {author} {\bibfnamefont {J.~J.}\
  \bibnamefont {van~den Berg}}, \bibinfo {author} {\bibfnamefont {I.~J.}\
  \bibnamefont {Vera-Marun}}, \bibinfo {author} {\bibfnamefont {P.~J.}\
  \bibnamefont {Zomer}}, \ and\ \bibinfo {author} {\bibfnamefont {B.~J.}\
  \bibnamefont {van Wees}},\ }\href {\doibase 10.1103/PhysRevB.90.235428}
  {\bibfield  {journal} {\bibinfo  {journal} {Phys. Rev. B}\ }\textbf {\bibinfo
  {volume} {90}},\ \bibinfo {pages} {235428} (\bibinfo {year}
  {2014})}\BibitemShut {NoStop}%
\bibitem [{\citenamefont {Krivorotov}\ \emph {et~al.}(2002)\citenamefont
  {Krivorotov}, \citenamefont {Gredig}, \citenamefont {Nikolaev}, \citenamefont
  {Goldman},\ and\ \citenamefont {Dahlberg}}]{krivorotov_role_2002}%
  \BibitemOpen
  \bibfield  {author} {\bibinfo {author} {\bibfnamefont {I.~N.}\ \bibnamefont
  {Krivorotov}}, \bibinfo {author} {\bibfnamefont {T.}~\bibnamefont {Gredig}},
  \bibinfo {author} {\bibfnamefont {K.~R.}\ \bibnamefont {Nikolaev}}, \bibinfo
  {author} {\bibfnamefont {A.~M.}\ \bibnamefont {Goldman}}, \ and\ \bibinfo
  {author} {\bibfnamefont {E.~D.}\ \bibnamefont {Dahlberg}},\ }\href {\doibase
  10.1103/PhysRevB.65.180406} {\bibfield  {journal} {\bibinfo  {journal} {Phys.
  Rev. B}\ }\textbf {\bibinfo {volume} {65}},\ \bibinfo {pages} {180406}
  (\bibinfo {year} {2002})}\BibitemShut {NoStop}%
\bibitem [{\citenamefont {Fert}\ and\ \citenamefont
  {Jaffrès}(2001)}]{fert_conditions_2001}%
  \BibitemOpen
  \bibfield  {author} {\bibinfo {author} {\bibfnamefont {A.}~\bibnamefont
  {Fert}}\ and\ \bibinfo {author} {\bibfnamefont {H.}~\bibnamefont
  {Jaffrès}},\ }\href {\doibase 10.1103/PhysRevB.64.184420} {\bibfield
  {journal} {\bibinfo  {journal} {Phys. Rev. B}\ }\textbf {\bibinfo {volume}
  {64}},\ \bibinfo {pages} {184420} (\bibinfo {year} {2001})}\BibitemShut
  {NoStop}%
\end{thebibliography}

%

\newpage

\begin{center}
 \textbf{\large Supplementary Information}
\end{center}

\section{sample preparation}
Graphene is mechanically exfoliated from a highly oriented pyrolytic graphite (HOPG) ZYA grade crystal on to  a pre-cleaned Si/SiO$_2$ substrate (300~nm thick SiO$_2$), where n$^{++}$ doped Si is used as a back gate electrode. Single layer graphene flakes were identified using an optical contrast. Ferromagnetic contacts are patterned via electron beam lithography on the PMMA (poly (methyl methacrylate))  coated graphene flake. Then, 0.8 nm of titanium (Ti) is deposited in two steps, each step of 0.4 nm of Ti deposition followed by in-situ oxidation by pure O$_{\text{2}}$ to form an oxide tunnel barrier to overcome the conductivity mismatch problem \cite{maassen_contacts}. On top of the oxide barrier we deposit 35 nm of cobalt for the spin selective contacts. To prevent oxidation of the ferromagnetic electrodes, the contacts are covered with 3 nm thick aluminum layer.

\section{Local charge Noise measurements and its magnetic field dependence}
For the noise measurements, we record 800 samples at a high sampling frequency (262 kHz) and measure the 1/f  noise in the frequency range of 25 Hz with the  resolution of 31.2 mHz. The final spectrum is recorded after performing the root mean square averaging over 20 FFT spectra.

\begin{figure}[!ht]
\includegraphics{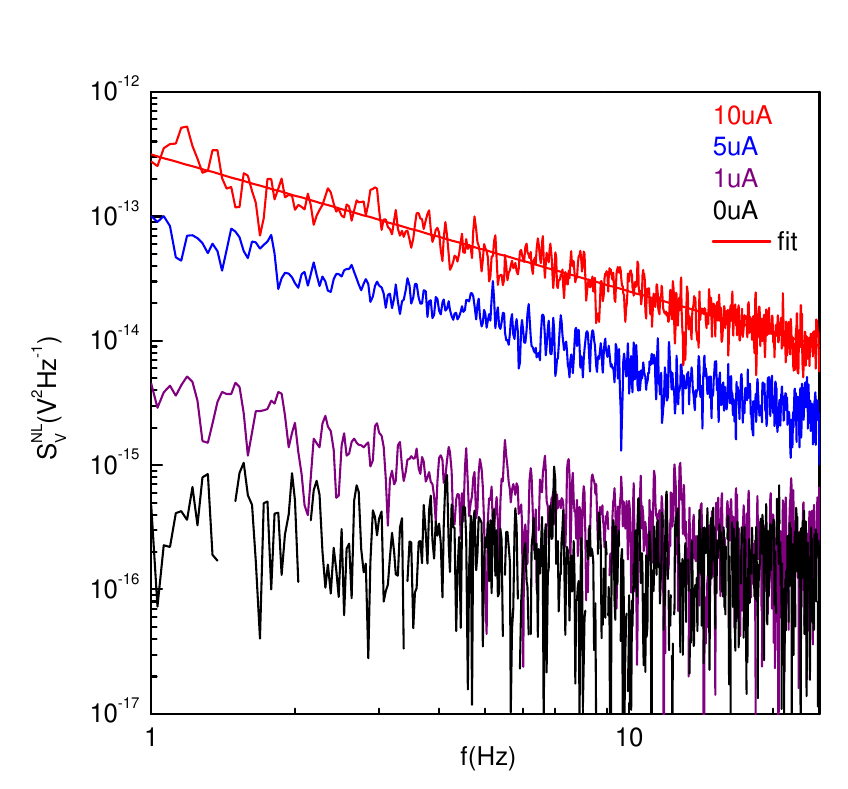}
\caption{\label{fig:flake noise} 1/f noise of the flake is measured in a local four terminal geometry (Fig.~1(a)) described in the main text.}
\end{figure}

We measure the charge 1/f noise associated with the flake in a local four probe measurement scheme, shown in Fig.~1(a) in the main text, using following equation\cite{hooge_1/f_1994}: 
\begin{equation}
 \frac{S_{\text{V}}}{V_{\text{local}}^2}=\frac{\gamma^{\text{c}}}{f^{a}}
 \label{eq:Hooge}
\end{equation}
where $S_{\text{V}}$ is the PSD of voltage fluctuations (units V$^2$/Hz), $\gamma^{\text{c}}$ is the noise magnitude i.e. the normalized Hooge parameter $\gamma^{\text{c}}_{\text{H}}$ with respect to the total number of charge carriers in the channel and characterizes the noise magnitude of the material, and $V_{\text{local}}$ is the average voltage drop across the sample. With the cross-correlation (XC) scheme, we are only sensitive to the noise from the conducting channel.  As expected, the noise increases with the bias current (Fig.~\ref{fig:flake noise}). We obtain $\gamma^{\text{c}} \sim 10^{-7}$ by fitting the spectrum with Eq.~\ref{eq:Hooge}. 
Here, we would also like to mention that the 1/f noise in Fig.~\ref{fig:flake noise} nicely scales with ${I_{\text{dc}}}^2$, implying that we are only sensitive to the 1/f noise fluctuations from the flake and the current source is not introducing the frequency dependent fluctuations from the contact through capacitive coupling. When the impedance of the current source becomes equivalent to the contact resistance at higher frequencies ($\sim \geq$ 10 MHz ) due to capacitive coupling, the noise in the injected spin current can come from the fluctuating contact resistance. In this case, the noise would increase at higher frequencies. On the other hand, we observe the opposite frequency dependence for the measured noise going down at higher frequencies complying with the 1/f noise behavior, and the noise is measured at very low frequencies where the impedance of the current source is almost constant and is much higher than the contact resistance, ruling out the effect of the contact noise on the measured signal.  

There could be local magnetoresistance (MR) contributions from the ferromagnetic contacts and the graphene flake present in our measurements. In order to rule out the flake MR contribution to the noise, we apply an out of plane magnetic field ($B_{\perp}$) with a dc current of 10~\textmu A is applied between the outer contacts and the noise is measured between the inner contacts. Similar measurements are performed for the contact magnetoresistance in a three terminal connection scheme where noise  at the graphene-tunnel barrier interface is measured. 
\begin{figure}
\includegraphics{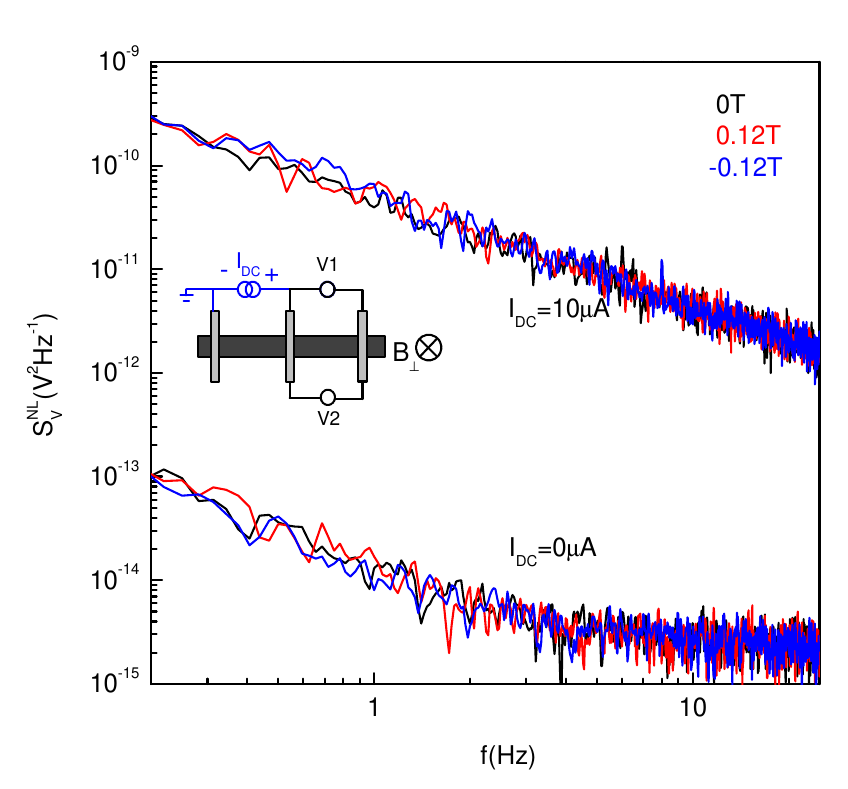}
\caption{\label{local MR contact} Noise associated with graphene-tunnel barrier interface measured via the XC method in a three terminal connection scheme (shown in the inset) for different values of $B_\perp$. }
\end{figure}
However, we do not observe a detectable change in the noise level at different magnetic fields for both the measurements (Fig.~\ref{local MR contact} and Fig.~\ref{local MR flake}). In this way we can discard the contribution of the MR coming from the graphene flake and the contacts to the observed magnetic field dependent non-local noise.
\begin{figure}
 \includegraphics{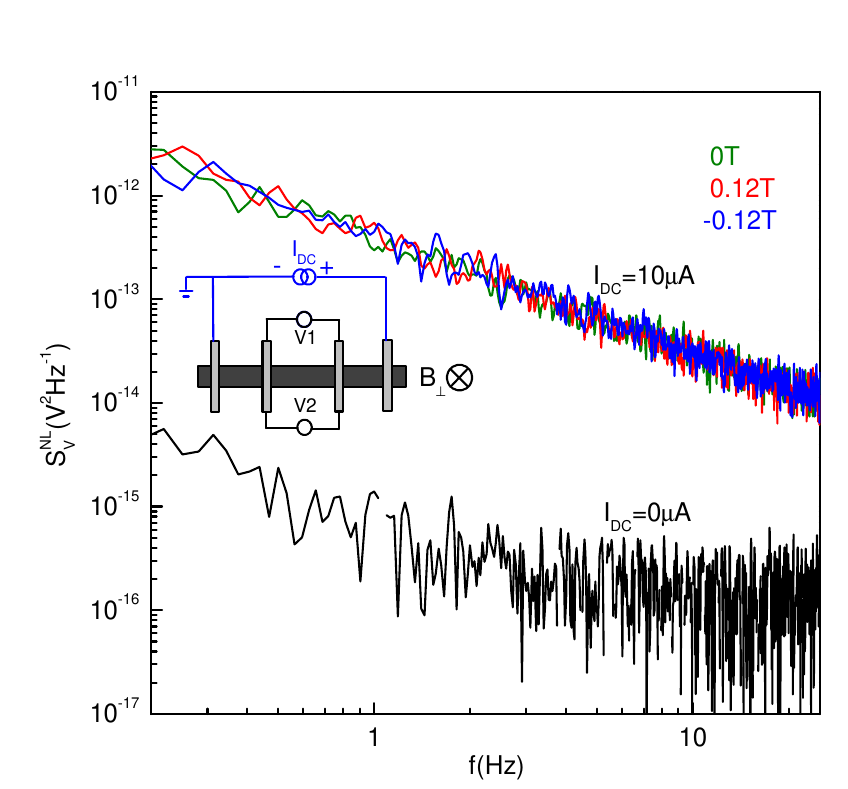}
 \caption{\label{local MR flake} Local noise of the graphene channel measured via the XC method as a function of $B_\perp$. The connection scheme is shown in the inset.}
\end{figure}

\section{Switching behaviour of the contacts: Spin valve measurement}
 In our spin-valve measurements (Fig. 2(a) of the main text), we observe an asymmetric switching of the cobalt FM contacts for the positive and the negative sweep of the in-plain magnetic field. This behavior has been observed before for spin transport in graphene \cite{berger_correlating_2015, guimaraes_spin_2014}. For cobalt contacts, if there is an  anti-ferromagnetic cobalt oxide layer formed on the side or top of a FM electrode, it can induce an exchange bias on the cobalt magnetization, which leads to a shift of the hysteresis loop and can cause different switching fields for positive and negative magnetic fields \cite{krivorotov_role_2002}. For the positive magnetic fields we observe the switching of all four electrodes. However, for the negative fields, two electrodes seem to switch imultaneously, and instead of four, we only detect three switches.
\section{Circuit Analysis}
\begin{figure}
 \includegraphics[scale=1]{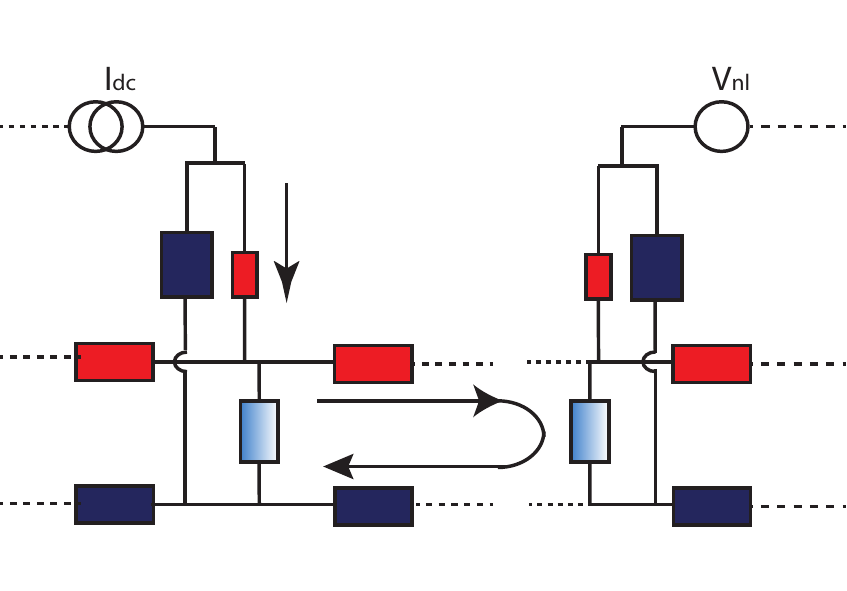}
 \caption{ \label{non local ckt}A two channel circuit model representing spin transport in  graphene in the non-local geometry. The horizontal resistor lines  represent the channel resistance for spin up (red) and spin down (blue) connected via a shunt spin relaxation resistance (blue-white). The vertical branches represent the ferromagnetic electrodes with spin up (small red resistor) and spin down resistance (big blue resistor). The transport channel is represented by a repetition of series and shunt resistance.}
\end{figure}

In order to identify the noise sources, contributing to the measured non-local noise,  we develop an elementary 2-channel resistor model as described in ref. \cite{fert_conditions_2001}, representing true spin and charge transport properties of the measured device (Fig.~\ref{non local ckt}).
A ferromagnetic injector/detector  is represented as a combination of two parallel resistors corresponding to spin-up and spin-down resistances, chosen in a way that they satisfy the condition for measured contact polarization $P=\frac{R^{\text{C}}_{\downarrow}-R^{\text{C}}_{\uparrow}}{R^{\text{C}}_{\downarrow}+R^{\text{C}}_{\uparrow}}$ and the contact resistance $R_{\text{C}}=\frac{R^{\text{C}}_{\downarrow}R^{\text{C}}_{\uparrow}}{R^{\text{C}}_{\downarrow}+R^{\text{C}}_{\uparrow}}$. Since graphene is non-magnetic, one can represent the spin resistance $R^{\text{s}}=R_{\text{sq}}\bigtriangleup{x}/W$ as a parallel combination 
of $R^{\text{s}}_{\uparrow}$ and $R^{\text{s}}_{\downarrow}$, where $R^{\text{s}}_{\uparrow}=R^{\text{s}}_{\downarrow}=2\times R^{\text{s}}$. Here $\bigtriangleup{x}$ is the length scale of graphene for which the channel and the relaxation resistance are defined. The model can be refined by taking smaller $\bigtriangleup{x}$. The spin-relaxation process in the circuit is represented by the relaxation resistor $R_{\uparrow \downarrow}=2R^{\text{s}}\times \lambda_{\text{s}}/\bigtriangleup{x}$.
For our simulation, we take the ratio $\lambda_{\text{s}}/\bigtriangleup{x} =3$, i.e. incorporating three relaxation resistors in one unit of $\lambda_{\text{s}}$. The thermal noise background is simulated by replacing each resistor by a noise-less resistor connected to the equivalent root mean squared (\textit{rms}) current noise source of $i_{noise}=\sqrt{\frac{4k_{\text{B}}T}{R \triangle f}}$ parallel to it at $\triangle$f=1 Hz.
The response of the equivalent current source is measured as a voltage difference between the detector pairs.  In this way, one can estimate the contribution from the relaxation shunt resistors, channel resistance, and the contacts separately. The total noise is the root mean squared sum of the contribution from all the circuit components. A major contribution of the simulated noise comes from the detector contacts, since the equivalent resistance of the circuit is dominated by the detector resistance and when there is no current flow in the circuit, one observes the equivalent thermal noise background. The value obtained from the simulation is $\sim 1.4 \times 10^{-16}$~V$^2$/Hz at room temperature and the thermal noise measured experimentally for our non-local circuit is $\sim 10^{-16}$~V$^2$/Hz, supporting the validity of our circuit model.

When a non-zero dc current flows through the graphene, 1/f charge noise is generated in addition to the background thermal noise. Coming back to our 2-channel resistor model, for  non-zero current, we know the amount of current (\textit{i}) flowing through each circuit element, which can be converted to the equivalent \textit{rms} noise current ${i_{noise}}^{1/f}=\sqrt{\gamma_{\text{H}}}*i$/f. Here we use $\gamma_{\text{H}}=\gamma^{\text{c}} \sim 10^{-7}$ and all the calculations are done at f~=~1 Hz. In a similar way, for the thermal noise simulation, we estimate the charge 1/f noise contribution at the non-local detector pair. The total 1/f noise ($\sim 6 \times 10^{-17}$~V$^2$/Hz)  due to charge 1/f noise is much lower than the noise we measure in the non-local geometry. For the thermal noise simulation, we can estimate the charge 1/f noise contribution from the channel resistors ($\sim 10^{-18}$~V$^2$/Hz ), relaxation resistors ($\sim 10^{-21}$~V$^2$/Hz ) and the contacts ($\sim 10^{-17}$~V$^2$/Hz) at the detector. The total 1/f noise ($\sim 6 \times 10^{-17}$~V$^2$/Hz) is again dominated by the detector 1/f noise. However, it is clear that the noise contribution due to charge 1/f noise is much lower than the noise we measure in the non-local geometry.  From the spin-dependent noise measurements at different spin accumulation, we experimentally obtain the proportionality constant $\gamma^{\text{s}} \sim 10^{-4}$ i.e.  $10^3$ times than the $\gamma^{\text{c}}$ for the charge noise. Using $\gamma_{\text{H}}=\gamma^{\text{s}}=10^{-4}$ we can simulate the noise level $\sim 10^{-14}$~V$^2$/Hz,  close to the observed noise level in our measurements.  We again confirm that it is only possible to see such a high noise non locally with a higher $\gamma_{\text{H}}$, which is not possible via the mechanism producing the charge 1/f noise. 

\section{Background Noise in non-local geometry}
\begin{figure}
\includegraphics{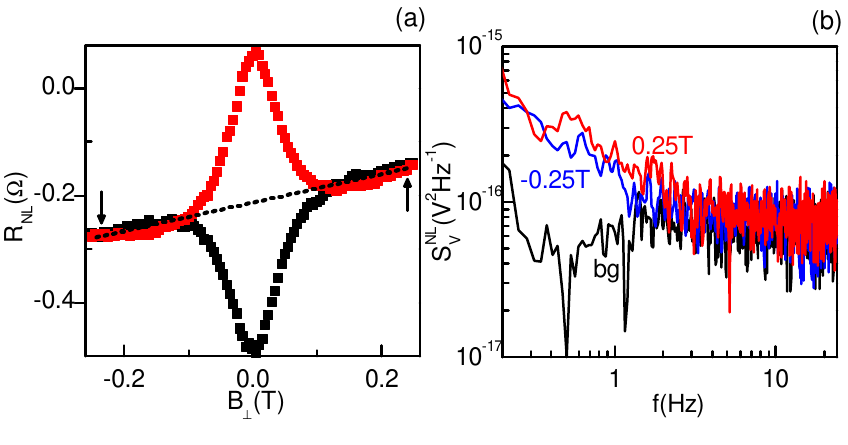}
 \caption{\label{hanle background} (a) Hanle precession  measurements for parallel [$\uparrow \uparrow$] and the anti-parallel [$\uparrow \downarrow$] magnetization of the inner injector and the detector electrodes. The background signal is denoted by black dash line (b) Non-local noise measurement for $B_{\perp}$ at -0.25 T and 0.25 T corresponding to the black vertical arrows in the Hanle curve.}
\end{figure}

In order to confirm that the magnetic field dependence of the measured noise is not originated by the non-local background, we measure the non-local noise at high positive and negative perpendicular magnetic fields ($B_{\perp} \sim$ 0.25 T and -0.25 T) where no spin accumulation is present (Fig.~\ref{hanle background}(a)). the non-local signal is different due to different background MR (dashed line in Fig.~\ref{hanle background}(a)). However, we do not observe any difference in the noise level for high positive and negative $B_{\perp}$, confirming that there is no detectable noise contribution from the non-local background signal(Fig.~\ref{hanle background}(b)). 
\section{spin-dependent noise: analytical expression}

We quantitatively analyze the analytical expression of the non-local spin signal  in order to figure out the dominant sources of spin sensitive noise.  The measured non-local voltage $\triangle V_{\text{NL}}=\triangle R_{NL}\times I$ for the spin-valve geometry is expressed by Eq.~\ref{non-local voltage}:
\begin{equation}
 \triangle V_{\text{NL}}=\frac{P^2IR_{\text{sq}}\lambda_{\text{s}}exp(-L/\lambda_{\text{s}})}{2W}
 \label{non-local voltage}
\end{equation}
Here $P$ is the contact spin polarization, $I$ is the current applied at the injector contact, $R_{\text{sq}}$ is the square resistance of graphene, $\lambda_{\text{s}}$ is the spin diffusion length in graphene, $L$ is the spacing between the injector and the detector contact and $W$ is the width of the transport channel. 

The fluctuations in $\triangle V_{\text{NL}}$ in time are represented by the correlation function:
\begin{equation}
R_V^{NL}(\tau)=<\triangle V_{\text{NL}}(t)*\triangle V_{\text{NL}}(t+\tau)>
\label{vnl correlation function}
\end{equation}
The noise associated with different parameters ($P, R_{\text{sq}}, I, \lambda_{\text{s}}$) can be written in form of a power spectrum ${S_{\text{V}}^{\text{NL}}}(f)$, which is the Fourier transform of ${R^{V}}_{NL}(\tau)$:
\begin{dmath}
 {S_{\text{V}}^{\text{NL}}}(f)\simeq\frac{{exp(-\frac{L}{{\lambda_{\text{s}}}})}^2}{4W^2}[{({{P}}^2{I}R_{\text{sq}})}^2\bigg(1+\frac{L}{\lambda_{\text{s}}}\bigg)^2S_{\lambda_{\text{s}}}(f)+{({{P}}^2{\lambda_{\text{s}}}R_{\text{sq}})}^2S_I(f)+
 {({{P}}^2{\lambda_{\text{s}}}{I})}^2S_{R_{\text{sq}}}(f)+{(2{P}{I}R_{\text{sq}}{\lambda_{\text{s}}})}^2S_{P}(f)]
\end{dmath}

This equation can be rewritten as
\begin{equation}
  {S_{\text{V}}^{\text{NL}}}(f)=A_{\lambda_{\text{s}}}S_{\lambda_{\text{s}}}(f)+A_IS_I(f)+A_{R_{\text{sq}}}S_{R_{\text{sq}}}(f)+A_PS_{P}(f)
 \label{power spectrum vnl}
\end{equation}
where $S_P$ is the due to the polarization fluctuations at the injector/detector electrode which is the Fourier transform of the auto correlation function for the time dependent polarization fluctuations i.e. $\mathcal{F}\langle P(t)P(t+\tau)\rangle$ , $S_{\lambda_{\text{s}}}$ is the noise associated with the spin transport i.e. spin relaxation noise ($\mathcal{F}\langle \lambda_{\text{s}}(t)\lambda_{\text{s}}(t+\tau)\rangle$), $S_I(f)$ is the noise from the external current source ($\mathcal{F}\langle I(t)I(t+\tau)\rangle$), $S_{R_{\text{sq}}}$ is the 1/f charge noise and the thermal noise from the channel ($\mathcal{F}\langle R_{\text{sq}}(t)R_{\text{sq}}(t+\tau)\rangle$). Here we take the assumption that the fluctuations in all four parameters ($P, R_{\text{sq}}, I, \lambda_{\text{s}}$) are uncorrelated. 
We can measure $S_I$ independently and $S_{R_{\text{sq}}}$ is the local 1/f noise (Fig.~\ref{fig:flake noise}). 
After removing the contribution of $S_I (\sim 10^{-23}$~V$^2$/Hz) and $S_{R_{\text{sq}}} (\sim 10^{-22}$~V$^2$/Hz) to ${S_{\text{V}}^{\text{NL}}}$, which are negligible compared to the observed noise ($\sim 10^{-14}$~V$^2$/Hz), the only dominant sources of noise in the measured non-local signal are the polarization fluctuations at the injector/detector electrodes and the fluctuations in the spin transport parameters ($\lambda_{\text{s}}=\sqrt{D_{\text{s}} \tau_{\text{s}}}$ ). However, assuming $R_{\text{sq}}$ and $\lambda_{\text{s}}$ uncorrelated is not strictly true. These quantities are correlated as $\lambda_{\text{s}}$ depends on the channel resistance  with $\lambda_{\text{s}}$ going down with the increase in the channel resistance, which would lead to different $\lambda_{\text{s}}$ dependence of the analytical expression i.e. Eq.~\ref{power spectrum vnl}.

\section{Non-local noise via cross correlation}

 \begin{figure}
 \includegraphics{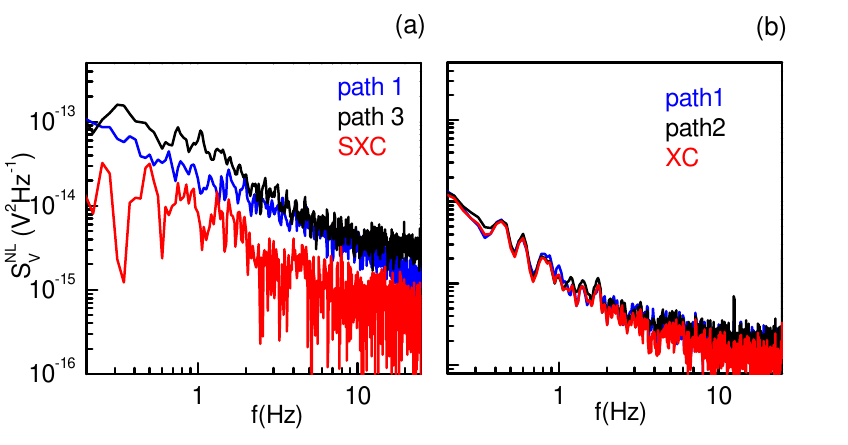}
 \caption{\label{fig:auto corr comparision}(a) Spin-dependent noise  measured from path 1 and path 3 and the SXC of the signals. (b) Spin-dependent noise  measured from path 1 and path 2 and the XC of the signals. Both the measurements are done following the connection scheme shown in  Fig.~1(b) in the main text.}
\end{figure}
  
 \begin{figure}[!ht]
\includegraphics{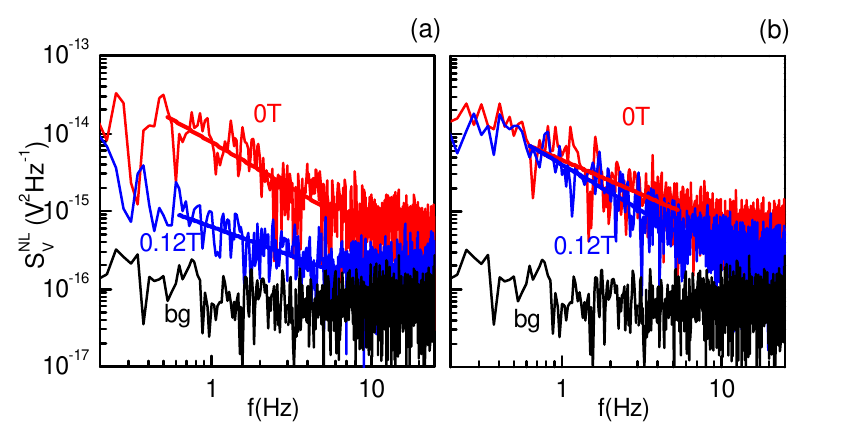}
 \caption{\label{fig:xcorr gate dependence} 
SXC noise (spin relaxation noise) measured at contact C4 at (a) $V_{\text{g}}$=0V and (b)$V_{\text{g}}$=-45V, for different magnetic fields. The spin relaxation noise at C4 for $V_{\text{g}}$=-45V is lower due to reduced spin transport parameters.}  
\end{figure}
 A measured non-local noise at the detector ($S_{\text{NL}}^V$) in the presence of spin-accumulation (spin-current) can be represented by Eq.~\ref{detector1 noise}:
 \begin{equation}
 S_{\text{NL}}^V={S_P}^{C1}+{S_P}^{C2}+{S_{\lambda_{\text{s}}}}^{C1-C2}+S_{bg}
  \label{detector1 noise}
 \end{equation}
Here $S_P^{C_i}$ is the contact polarization noise, $S_{\lambda_{\text{s}}}$ is the noise due to spin accumulation (relaxation) between contacts C1-C2 and $S_{bg}$ is the electronic 
noise contribution due to residual charge current flowing in the non-local circuit and the thermal noise background. $S_{bg}$ does not carry any spin-dependent information. On applying a high magnetic field ${B_{\perp}} \sim$0.1~T, perpendicular to  the device plane, one can suppress the spin transport and the 
measured non-local noise contribution at high $B_{\perp}$ can come from the background charge 1/f noise ($S_{bg}$) due to non-homogeneous charge current distribution in the non-local regime. In this way the spin-dependent noise (polarization and spin accumulation) can be separated from the total noise. 

Polarization fluctuations in each contact are independent from each other and one can filter the polarization noise from the spin current noise by
using the spatial XC method. We measure the non-local noise via XC scheme as shown in Fig.~1(b) in the main text. Simultaneously we also record the 
single channel noise for contact pair C3-C5 (path 1) and C4-C5 (path 3). Single channel noise includes the polarization noise contribution of the contact pair on top of the spin relaxation noise between the contacts. These contributions can be summarized in following equations:
\begin{equation}
 {S_{NL}}^{channel1}={S_P}^{C3}+{S_P}^{C5}+{S_{\lambda_{\text{s}}}}^{C3-C5} \label{autocorr ch1}
\end{equation}
\begin{equation}
 {S_{NL}}^{channel2}={S_P}^{C4}+{S_P}^{C5}+{S_{\lambda_{\text{s}}}}^{C4-C5} \label{autocorr ch2}
\end{equation}
  \begin{equation}
  {S_{xcorr}}^{C3-C5\otimes C4-C5}={S_{\lambda_{\text{s}}}}^{C3-C5}+{S_P}^{C5} \label{xcorr ch1-2}
 \end{equation}
Note that we have not included the background noise contribution term here as it can be estimated separately at ${B_{\perp}} \sim$~0.1~T 
via the procedure described above and the final equations can be rewritten without the background contribution. 
On the other hand, the spatial cross correlation of $V_{C3-C5}$ and $V_{C4-C5}$  will have total noise contribution ${S_{xcorr}}^{C3-C5\otimes C4-C5}$ only from the outer detector C5 (${S_P}^{C5}$) and the spin relaxation noise ${S_{\lambda_{\text{s}}}}^{C4-C5}$.

\begin{figure}[!ht]
\includegraphics{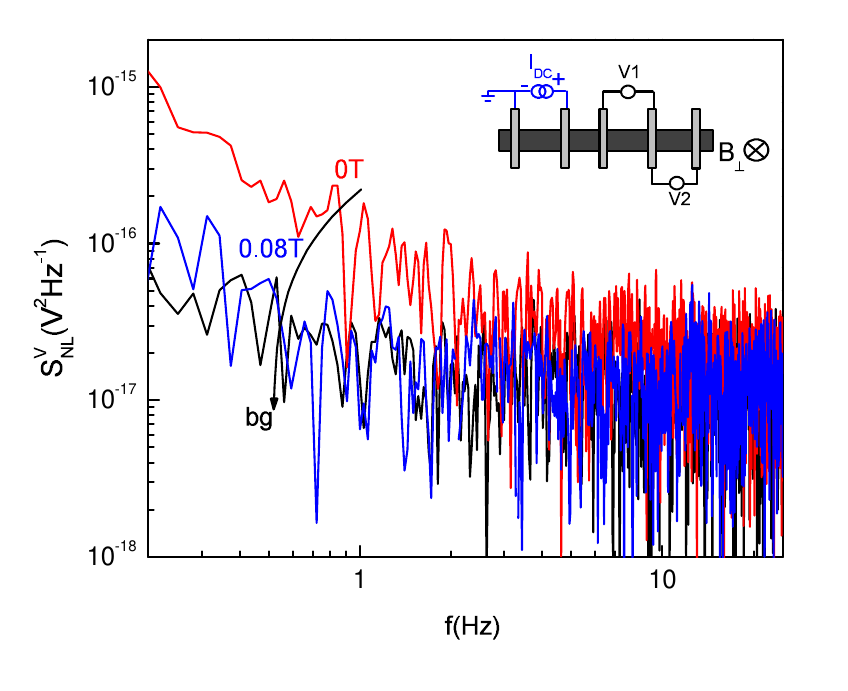}
 \caption{\label{fig:polarization noise} 
Contact polarization noise measured for contact C4 at B~=~0~T (red) and at B~=~80~mT. For the current injection at C1-C2, the spin-dependent noise between contacts C3-C4 and C4-C5 are cross correlated and only the noise at C4 is measured. For $B_{\perp} \sim$~80~mT, the polarization noise is reduced to the background noise in the absence of the spin accumulation underneath the contact.The connection scheme is shown in the inset.}  
\end{figure}

The advantage of spatial cross correlation over the regular cross correlation method is demonstrated in Fig.~\ref{fig:auto corr comparision}(a). For the regular XC measurement, we do not observe any difference between the single channel noise and the noise computed via XC method Fig.~\ref{fig:auto corr comparision}(b). The reason for this is the contribution from the contact leads and the external electronics is much lower than the measured noise level between the contacts C3 and C5. Therefore, the uncorrelated signals do not affect the measurement. On the other hand, by using the spatial cross correlation method one can eliminate the polarization noise contribution from different detectors and the method allows to see the noise contribution which is shared between the detector pairs C3-C5 and C4-C5 i.e. spin relaxation noise between contacts C4 and C5 ($\sim$ 5$\times$10$^{-15}$ V$^2$Hz$^{-1}$), without the contribution of the polarization noise from C4. Using $S_{\lambda_{\text{s}}} \propto  {\mu_{\text{s}}}^2$ relation, we can extrapolate the spin relaxation noise at contact C3 (path 1), which should be approximately four times higher than the spin relaxation noise at C4 for $\lambda_{\text{s}}$=1.5 $\mu$m i.e. ($\sim$10$^{-14}$ V$^2$Hz$^{-1}$).  From Fig.~\ref{fig:auto corr comparision}(a), we can say that the extrapolated spin relaxation noise  is almost equal to the spin-dependent noise measured via path 1 ($\sim$ 1.4 $\times$10$^{-15}$ V$^2$Hz$^{-1}$ at f = 1 Hz), leading to the conclusion that the spin-dependent noise at C3 is dominated by the spin relaxation noise in graphene.

We also measure the contact polarization noise separately by cross correlating the noise measured from the detector pairs C3-C4 and C4-C5, while C1 and C2 are the current injectors. Here only the noise from contact C4 is measured for different values of $B_{\perp}$ ( spin accumulation) underneath the contact. We clearly see the spin-dependent noise (contact polarization noise in this case) is reduced to the background noise at $B_{\perp} \sim$~ 80~mT, where spin accumulation is suppressed (Fig.~\ref{fig:polarization noise}). The polarization noise is $\sim$ 10$^{-16}$ V$^2$Hz$^{-1}$ (at 1~Hz), which is negligible compared to the measured spin relaxation noise i.e. $\sim$ 10$^{-14}$ V$^2$Hz$^{-1}$.

\end{document}